\documentclass[journal,web]{article}
\usepackage[utf8]{inputenc}

\usepackage{amsmath,amsfonts,amssymb}
\usepackage{graphicx}
\usepackage{xcolor}
\graphicspath{{./graphics/}}

\usepackage[margin=2cm]{geometry}
\usepackage{multicol}
\usepackage{caption}
\usepackage{float} 
\usepackage{cite}

\usepackage{tikz}
\usetikzlibrary{calc, svg.path}

\usepackage{pgfplots}
\pgfplotsset{compat=newest}

\usepackage{dblfloatfix} 
\usepackage{pbox} 

\usepackage{nicefrac}

\usepackage{bbm}
\newcommand{\IN}{\mathbbm{N}}		
\newcommand{\IR}{\mathbbm{R}}		
\newcommand{\IC}{\mathbbm{C}}		

\renewcommand{\epsilon}{\varepsilon}					
\renewcommand{\theta}{\vartheta}                  	    
\renewcommand{\phi}{\varphi}							

\newcommand{\norm}[1]{\left\lVert #1 \right\rVert}		

\newcommand{\mat}[1]{\boldsymbol{#1}}

\usepackage[binary-units = true,exponent-product = \cdot]{siunitx}
\def\num#1{\numx#1}
\def\numx#1e#2#3{{#1}\cdot10^{#2#3}}
\DeclareSIUnit{\mup}{\text{$\mu_0$}}
\DeclareSIUnit{\sample}{S}
\DeclareSIUnit{\mT}{\milli\tesla}
\DeclareSIUnit{\proceduredefinedunit}{p.d.u.}
\DeclareSIUnit{\kHz}{\kilo\hertz}
\DeclareSIUnit{\MHz}{\mega\hertz}
\DeclareSIUnit{\Tpm}{\tesla\per\metre}
\DeclareSIUnit{\mul}{\micro\litre}
\DeclareSIUnit{\mgml}{\milli\gram/\milli\litre}
\DeclareSIUnit{\cm}{\centi\metre}
\DeclareSIUnit{\mm}{\milli\metre}
\DeclareSIUnit{\mum}{\micro\metre}
\DeclareSIUnit{\mumol}{\micro\mol}
\DeclareSIUnit{\degC}{\celsius}

\definecolor{ibilight}{RGB}{193,216,237}
\definecolor{ibidark}{RGB}{0,73,146}	
\definecolor{uke2}{RGB}{170,156,143} 	
\definecolor{uke3}{RGB}{87,87,86}		
\definecolor{ukesec1}{RGB}{255,223,0}	
\definecolor{ukesec2}{RGB}{239,123,5}	
\definecolor{ukesec3}{RGB}{104,195,205}	
\definecolor{ukesec4}{RGB}{138,189,36}	
\definecolor{ukesec5}{RGB}{178,34,41}	
\definecolor{tuhh}{RGB}{45,198,214}     

\definecolor{ibidarkBG}{RGB}{227,229,242}   
\definecolor{uke2BG}{RGB}{233,228,225} 	    
\definecolor{uke3BG}{RGB}{230,231,232}	    
\definecolor{ukesec1BG}{RGB}{255,243,190}   
\definecolor{ukesec2BG}{RGB}{254,232,212}   
\definecolor{ukesec3BG}{RGB}{222,241,241}   
\definecolor{ukesec4BG}{RGB}{233,243,222}   
\usepackage[nolist,nohyperlinks]{acronym}

\begin{acronym}[ECU]

\acro{AWMPS}[AWMPS]{arbitrary waveform magnetic particle spectrometer}
\acro{ADC}[ADC]{analog-to-digital converter}
\acro{DAC}[DAC]{digital-to-analog converter}

\acro{CT}[CT]{Computed Tomography}

\acro{FFL}[FFL]{field-free-line}
\acro{FFP}[FFP]{field-free-point}
\acro{FFR}[FFR]{field-free-region}
\acro{FFT}[FFT]{fast Fourier transform}
\acro{FOV}[FOV]{field of view}

\acro{GUI}[GUI]{graphic user interface}

\acro{ISI}[ISI]{integrated signal intensity}
\acro{ICs}[ICs]{integrated circuits}

\acro{LFR}[LFR]{low-field-region}
\acro{LNA}[LNA]{low noise amplifier}

\acro{MPI}[MPI]{Magnetic Particle Imaging}

\acro{PSF}[PSF]{point spread function}
\acro{PNS}[PNS]{peripheral nerve stimulation}

\acro{SNR}[SNR]{signal-to-noise ratio}
\acro{SAR}[SAR]{specific absorption rate}
\acro{SPIONs}[SPIONs]{superparamagnetic iron oxide nanoparticles}

\acro{TF}[TF]{transfer function}

\end{acronym}

\usepackage{scalerel}
\definecolor{orcidlogocol}{HTML}{A6CE39}
\tikzset{
  orcidlogo/.pic={
    \fill[orcidlogocol] svg{M256,128c0,70.7-57.3,128-128,128C57.3,256,0,198.7,0,128C0,57.3,57.3,0,128,0C198.7,0,256,57.3,256,128z};
    \fill[white] svg{M86.3,186.2H70.9V79.1h15.4v48.4V186.2z}
                 svg{M108.9,79.1h41.6c39.6,0,57,28.3,57,53.6c0,27.5-21.5,53.6-56.8,53.6h-41.8V79.1zM124.3,172.4h24.5c34.9,0,42.9-26.5,42.9-39.7c0-21.5-13.7-39.7-43.7-39.7h-23.7V172.4z}
                 svg{M88.7,56.8c0,5.5-4.5,10.1-10.1,10.1c-5.6,0-10.1-4.6-10.1-10.1c0-5.6,4.5-10.1,10.1-10.1C84.2,46.7,88.7,51.3,88.7,56.8z};
  }
}
\newcommand\orcidicon[1]{\href{https://orcid.org/#1}{\raisebox{3pt}{\hspace{2pt}\mbox{\scalerel*{
    \begin{tikzpicture}[yscale=-1,transform shape]
    \pic[]{orcidlogo};
    \end{tikzpicture}
}{-}}}}} 

\usepackage[hidelinks]{hyperref} 

\newcommand\blfootnote[1]{%
  \begingroup
  \renewcommand\thefootnote{}\footnote{#1}%
  \addtocounter{footnote}{-1}%
  \endgroup
}

\title{\fontsize{28}{28}\selectfont {System Matrix based Reconstruction for \\ Pulsed Sequences in Magnetic Particle Imaging}}

\date{August 2021}

\author{
    Fabian Mohn\orcidicon{0000-0002-9151-9929}, 
    Tobias Knopp\orcidicon{0000-0002-1589-8517}, 
    Marija Boberg\orcidicon{0000-0003-3419-7481}, 
    Florian Thieben\orcidicon{0000-0002-2890-5288}, 
    Patryk Szwargulski\orcidicon{0000-0003-2563-9006} \\ and
    Matthias Graeser\orcidicon{0000-0003-1472-5988}
}

\begin{document}
\fontsize{10}{11.5}\selectfont

\maketitle


\begin{abstract} 
\noindent Improving resolution and sensitivity will widen possible medical applications of magnetic particle imaging. 
Pulsed excitation promises such benefits, at the cost of more complex hardware solutions and restrictions on drive field amplitude and frequency.
State-of-the-art systems utilize a sinusoidal excitation to drive superparamagnetic nanoparticles into the non-linear part of their magnetization curve, which creates a spectrum with a clear separation of direct feed-through and higher harmonics caused by the particles response.
One challenge for rectangular excitation is the discrimination of particle and excitation signals, both broad-band. 
Another is the drive-field sequence itself, as particles that are not placed at the same spatial position, may react simultaneously and are not separable by their signal phase or shape. 
To overcome this potential loss of information in spatial encoding for high amplitudes, a superposition of shifting fields and drive-field rotations is proposed in this work. 
Upon close view, a system matrix approach is capable to maintain resolution, independent of the sequence, if the response to pulsed sequences still encodes information within the phase.
Data from an Arbitrary Waveform Magnetic Particle Spectrometer with offsets in two spatial dimensions is measured and calibrated to guarantee device independence. 
Multiple sequence types and waveforms are compared, based on frequency space image reconstruction from emulated signals, that are derived from measured particle responses.
A resolution of 1.0$\,$mT (0.8$\,$mm for a gradient of (-1.25,\,-1.25,\,2.5)$\,$Tm$^{\textup{-1}}$) in x- and y-direction was achieved and a superior sensitivity for pulsed sequences was detected on the basis of reference phantoms.


\end{abstract}

\blfootnote{This work was supported by the German Research Foundation (DFG, grant numbers GR 5287/2-1, KN 1108/7-1), the Forschungszentrum Medizintechnik Hamburg (grant number 01fmthh2018). The Fraunhofer IMTE is supported by the EU (EFRE) and the State Schleswig-Holstein, Germany (Project: IMTE – Grant: 124 20 002 / LPW-E1.1.1/1536). \\
F. Mohn, T. Knopp, M. Boberg, F. Thieben, P. Szwargulski, and M. Graeser are with the Section for Biomedical Imaging, University Medical Center Hamburg-Eppendorf, 20251 Hamburg, Germany and with the Institute for Biomedical Imaging, Hamburg University of Technology, 21073 Hamburg, Germany (e-mail: fabian.mohn@tuhh.de).\\
M. Graeser is also with the  Fraunhofer Research Institute for Individualized and Cell-based Medicine and the Institute for Medical Engineering, University of Lübeck, 23562 Lübeck, Germany}

\vspace{0.2cm}
\begin{multicols}{2}

\section{Introduction}
\label{sec:intro}
In \ac{MPI} the spatial distribution of \ac{SPIONs} is determined by a superposition of a static gradient field and one or several oscillating excitation fields \cite{Gleich2005Nature}. 
The static gradient field, called selection field, generates a \ac{LFR} in its center, that includes a \ac{FFR}, which could either be a \ac{FFP} or \ac{FFL}, depending on its shape. On the one hand, the oscillating excitation field drives the \ac{SPIONs} through their magnetization curve, causing higher harmonics due to their nonlinear characteristic. On the other hand, it drives the \ac{LFR} through the imaging volume and creates a specific trajectory.
Most scanner topologies use narrow-band sinusoidal signal shapes to excite the tracer material (see Table II in \cite{Panagiotopoulos2015} for an overview). Due to the superposition of the oscillating fields and the selection field, these encoding schemes cause the \ac{SPIONs} in the \ac{LFR} to respond at a specific point in time with their maximal amplitude. In frequency domain this results in a spectral fingerprint depending on space, which can be used to reconstruct the image by solving a linear system of equations \cite{Gleich2005Nature,Knopp2010PhysMedBio}.
Another advantage of narrow-band excitation is the discrimination of the frequency space
in the two domains, the narrow-band excitation band and the higher receive band, which contains the harmonics caused by the particles non-linearity. Due to this discrimination, a separation between the strong feed-through of the excitation field and the by $10^{-6}$ to $10^{-10}$ lower particle signal can be achieved, using resonant passive filtering \cite{Graeser2013}. 
Current developments show that such encoding schemes can provide sub-millimeter resolution \cite{Vogel2019Micro}, more than 46 volumes per second time resolution \cite{Weizenecker2009} and pico-gram sensitivity \cite{Graeser_2020}. Developments in instrumentation now reach for clinical scale \cite{graeser2019Head,Mason2017, Rahmer2018} to address specific needs, which are currently only partly addressed by conventional imaging systems. Possible medical applications reach from catheter imaging \cite{Haegele2013b} in digital subtraction angiography, over stent quantification \cite{Wegner2019}, stroke imaging \cite{Szwargulski2020ACS, GraeserHeadCoil2020,Ludewig2017Stroke} and many more. Using multi-contrast image reconstruction \cite{rahmer2015first}, it is also possible to distinguish between different particle systems \cite{Shasha2019} or physical parameters in the vicinity of the particle system like temperature \cite{stehning2016simultaneous}, viscosity\cite{M_ddel_2018}, or binding state \cite{moddel2020estimating,Viereck2017}.
With sinusoidal excitation, large single-core particles show a strong relaxation behaviour, which broadens the \ac{PSF} and reduces the signal response, therefore reducing sensitivity and resolution \cite{Tay_relaxation_wall_2017}.
Recently, Tay et al. proposed a rectangular excitation and showed that it has the potential to improve the achievable resolution by using large particles in combination with a new reconstruction approach. \cite{tay2019pulsed}. 
They showed that the effect of broadening the \ac{PSF} can be significantly reduced under certain conditions, using rectangular excitation with small amplitudes \cite{tay2019pulsed}. 
In idealized rectangular excitation sequences, a \ac{LFR} would jump between two resting points in space, causing all particles in between to react simultaneously. Tay et al. proposed a reconstruction scheme, that integrates the receive signal, which in turn encodes the magnetic moment $\mat{m}$ of the area between the \ac{LFR} locations before and after the pulse, as long as $\mat{m}$ reaches a steady state within the resting time between pulses.
Consequently, this approach requires low field amplitudes (\SIrange{1}{3}{\mT}), such that the \ac{LFR} remains within a single voxel and no loss in spatial resolution in the excitation direction is induced. Otherwise, the information of the particle distribution between two \ac{LFR} positions would be lost. 
Furthermore, a low excitation frequency is required in order to let the particles fully relax to ensure their steady state \cite{tay2019pulsed}. 

In this paper, multiple sequences based on different excitation waveforms and amplitudes are compared in system matrix phantom reconstructions. A sequence for rectangular excitation shapes is proposed, that uses shifts and rotations simultaneously for better spatial encoding. In this proposed sequence, the rectangular excitation is rotated and shifted orthogonal to the excitation. 
Similar to a radon sampling scheme, this allows to reconstruct measured data with large amplitudes and an isotropic sampling trajectory in 2D, even if the phase information between the resting points is lost.
If particles relax fast enough to follow the slew rate of the drive field, their phases within the signals become distinguishable which allows a system matrix approach to reconstruct images at high resolution.
For pulsed sequences, the results show improved sensitivity while the proposed sequence has uniform and high resolution with short acquisition times.

\section{Motivation and Concept}
\label{sec:motivation}

All signals and reconstructions are emulated, meaning they are based on measurements of an \ac{AWMPS}\cite{tay2016, Pantke2019MultiFreq,Top2020AWMPS}, which are processed to resemble a specific particle response for a defined sequence.
By exciting \ac{SPIONs}, superimposed by a range of different DC-offsets in two spatial dimensions, their response is mapped on a 2D grid. This resembles the response to an overlaying 2D gradient field \cite{vonGladiss2017}. This approach has the benefit to deliver reliable results for a first evaluation, without the time consuming process of building an actual system that can produce the required fields on a large scale.
Following conditions should be met to reproduce the data:
\paragraph{Hardware} A device that is capable of producing arbitrary waveform excitations, superimposed by two orthogonal spatial DC-offsets. See Section \ref{sec:AWMPS} for details on the \ac{AWMPS}, which is used to measure a hybrid system matrix with 1D excitation and 2D offset fields \cite{vonGladiss2017}. This raw dataset is later processed to generate the sequence specific particle response.
\paragraph{Transfer function correction} To be able to compare the results with other imaging systems, the data needs to be handled in a comparable physical variable, in this case the domain of the magnetic moment $\mat{m}$ [\si{\A\m^2}], which is achieved by correcting the received signals with the \ac{TF} of the system \cite{Gladiss2020a}. In this domain, the signal is independent of the receiver and individual properties of a given system, like receive coils or amplification \cite{Knopp2010_modelbased}. This allows cross-platform comparison of the \ac{SPIONs} response to a given excitation.
\paragraph{Sequence noise} To avoid correlation of intrinsic measurement noise by reusing a subset of the same dataset within the emulated sequence, the signal needs to be overlaid with dominant, digitally generated noise $\tilde{u}^s_\textup{noise}(t)$. 
\paragraph{Reconstruction noise} A model to accurately represent receive chain noise of a scaled \ac{MPI} system, consisting of resonant coils, \ac{LNA} and \ac{ADC}, based on reference measurements from a \SI{40}{mm} receive coil, named $\tilde{u}^r_\textup{noise}(t)$ \cite{Graeser2017SR}. Therefore, calculated images are comparable to those of a small scanner system.
\paragraph{Phantoms} In spite of correcting data by a \ac{TF}, the measured data from the \ac{AWMPS} has a high \ac{SNR}, due to the close proximity of the receive coils to the tracer and the large sample compared to a voxel volume of a system matrix calibration. While the close proximity is corrected by the \ac{TF}, a realistic model for scaling the measured iron mass to the emulated voxel grid is needed. 
\paragraph{Independent system matrices} Two sets of system matrices need to be acquired to avoid inverse crime, when the phantom spectrum stems from the identical dataset as the system matrix used for reconstruction. For each sequence that is investigated in this paper, two independent system matrices on different grids are calculated, which are based on separately measured datasets, to guarantee a realistic reconstruction.
\paragraph{Comparability across sequences} In a final condition, the total sequence measurement time is chosen as the common criterion to compare sequences of different design. \\

\noindent Following these criteria, the images reconstructed by this method represent realistic images, achievable by a well designed pulsed \ac{MPI} system.

\section{Methods}
\label{sec:methods}

Different sequences are generated on various sets of raw data, to compare and identify the effects that excitation waveform, amplitude or sampling trajectory have on the image \ac{SNR} and resolution. This includes pulsed and sinusoidal excitation, as well as low and high excitation amplitudes. Individual sequences and their sampling trajectories are explained in Section \ref{sec:sequence}.  
Instead of depending on a narrow \ac{PSF} for high resolution, the proposed method in this work uses higher excitation field strengths above \SI{10}{\mT}, as this benefits not only the sensitivity, but also reduces acquisition time in future imaging systems due to the capability of using less averages.
To highlight the advantages and limitations between the different methods, the Cartesian sequence proposed in \cite{tay2019pulsed} is compared with the proposed method in this work. 
For a detailed comparison of the benefits of pulsed excitation, the proposed shift-radial sequence is run with both, rectangular excitation as well as sinusoidal excitation. 
A radial sine sequence is also included, as proposed by Knopp et al. in \cite{knopp2008trajectory}, which does not use the shift orthogonal to the excitation direction. The system matrix reconstruction approach is chosen for all sequences \cite{Knopp2010PhysMedBio}.
For an overview of the image reconstruction pipeline in this study, the process is depicted in Fig.~\ref{fig:AWMPS} (c).

\subsection{Arbitrary Waveform Magnetic Particle Spectrometer}
\label{sec:AWMPS}

\begin{figure}[H] 
    \centering
    \includegraphics[width=1.0\linewidth]{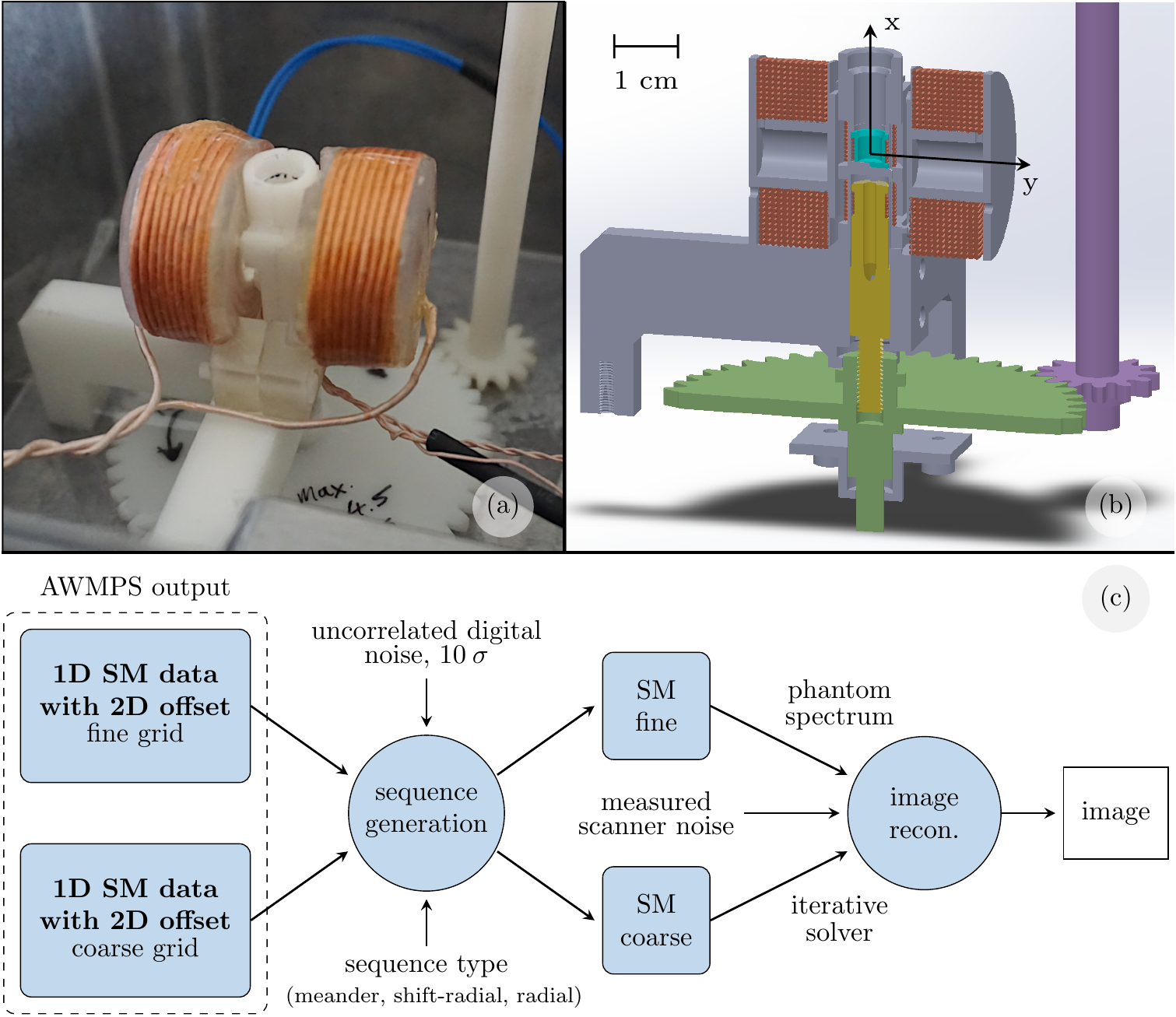}
    \caption{\textbf{Sectional view of the \acs{AWMPS} and flowchart of the entire process chain from raw data to image.} Using gradiometric receive coils, the receive signal can be isolated by suppressing direct excitation feed-through. Fine-tuning is achieved by turning the gear-wheel. A Helmholtz-coil creates an additional DC-offset field in another spatial dimension ($y$-direction).
    The general process of the methodology is shown in the block diagram.}
    \label{fig:AWMPS}
\end{figure}

A non-resonant \ac{AWMPS} with two transmit coils was built to perform measurements that contain the one-dimensional excitation, superimposed by DC-offsets in two orthogonal spatial dimensions. The main transmit coil combines excitation signal and DC-offset in $x$-direction within a single cylindrical coil, whereas a second transmit coil in Helmholtz configuration is responsible for the orthogonal DC-offset in $y$-direction with up to \SI{\pm 50}{\mT}. The \ac{AWMPS} may measure any waveform up to \SI{45}{\mT} in amplitude and is limited for arbitrary pulse shapes by a final slew-rate due to load and amplifier characteristics, around \SI{10}{\mT\per\us}. Fig. \ref{fig:AWMPS} shows a picture and the cross-section of the design.
The excitation and $x$-offset coil is connected to a 4-quadrant-amplifier (\textit{Dr. Hubert GmbH, Bochum, Germany}). To avoid direct feed-through, two receive coils are arranged in opposite orientation for decoupling of the receive path from the transmission line. The position of one coil is adjustable by a gear wheel for fine-tuning of the feed-through cancellation. The signal is then amplified by a custom-build \ac{LNA}. 
The data acquisition card \textit{STEMLab 125-14, Red Pitaya} is used for signal generation on two transmit channels and records the receive signal, as well as the amplifiers reference current monitor for control purposes.

\subsection{Measurements}
\label{sec:measurements}

\begin{figure}[H] 
    \centering
    \includegraphics[width=1.0\linewidth]{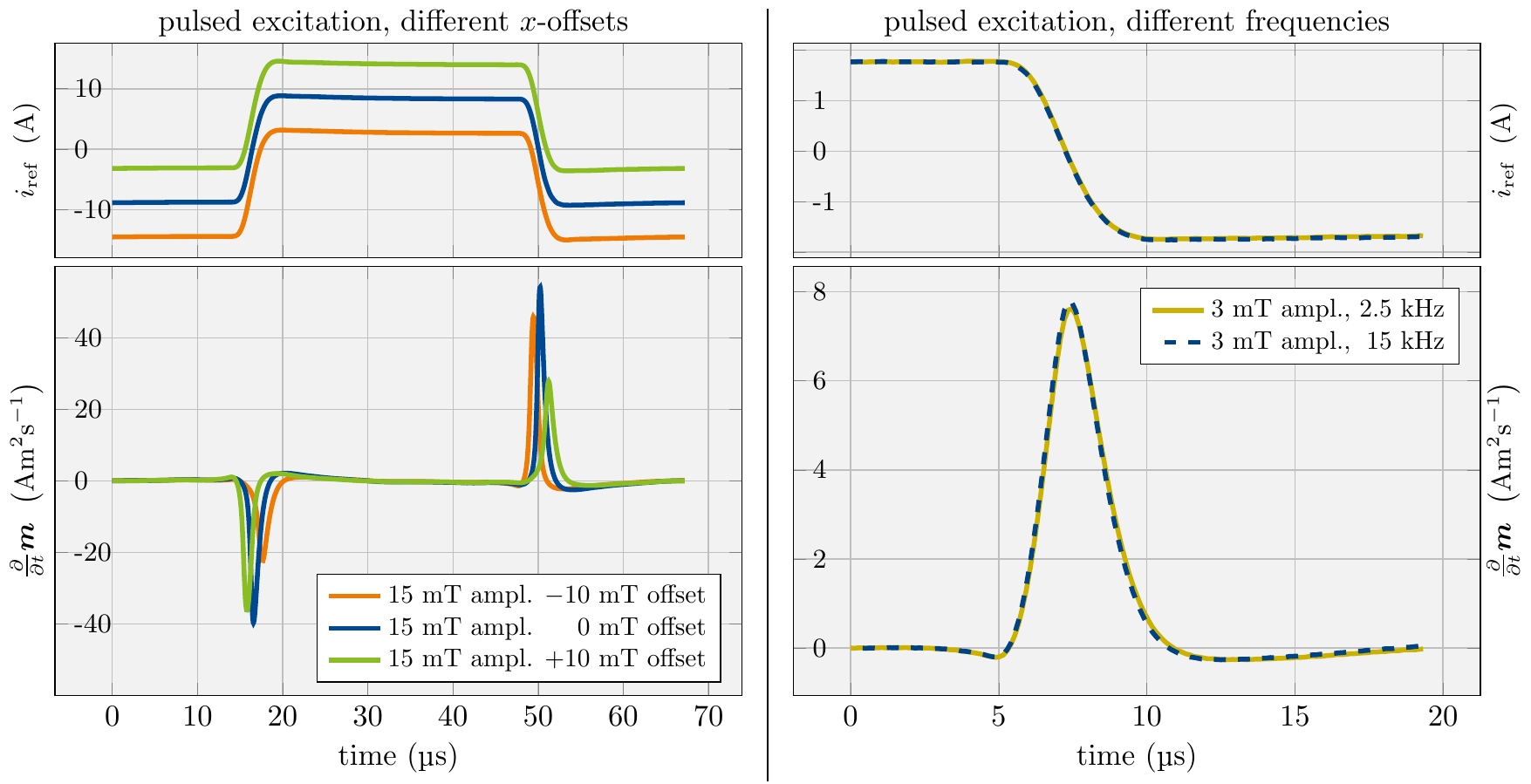}
    \caption{\textbf{Pulsed excitation measurements with a $\SI{20}{\micro\liter}$ sample of Perimag $\SI{17}{\ug\of{Fe}\per\micro\liter}$.} To the left, three offsets in excitation direction are plotted (\SI{14.88}{\kHz}), the reference channel from the current monitor is on the top, the \acs{TF} corrected receive signal on the bottom.
    Plots on the right show identical relaxation behavior of two overlaid signals at \SI{2.5}{\kHz} and \SI{14.88}{\kHz}. The slew rate of both frequencies is identical, only the hold time differs. Both signal curves are coinciding for Perimag.}
    \label{fig:plot_freq}
\end{figure}

All measurements in this work are based on a multi-core tracer, Perimag (\textit{micromod Partikeltechnologie GmbH, Rostock, Germany}) with an undiluted iron concentration $c_\textup{meas}$ of \SI{17}{\mg\of{Fe} \per \milli\liter} (\SI{304.4}{\mmol \per \liter}). The delta sample is filled with $V_\textup{meas} = \SI{20}{\ul}$ of undiluted tracer to guarantee high measurement \ac{SNR} at low amplitudes. The excitation frequency is chosen to be \SI{14.88}{\kHz}, to ensure steep flanks of the rectangle excitation and avoid slew rate artifacts on the edges while keeping a good reception for induction sensors.
This differs from the choice of \SI{2.5}{\kHz} in \cite{tay2019pulsed}, however the rise time $t_\textup{r}$ of the pulsed excitation (square wave, $t_\textup{r}=\SI{3}{\us}$) lies in a similar region (\SIrange{2}{5}{\us}). On the right side in Fig. \ref{fig:plot_freq} these two frequencies are compared, to confirm the quick and identical relaxation behaviour of the used tracer for both frequencies.
We note that Perimag is a tracer with low relaxation times, in contrast to the long relaxations times of the tracers used in \cite{tay2019pulsed}. Therefore, our frequency choice is only valid for fast relaxing particles and not a general choice for larger core \ac{SPIONs}.
Instead of averaging $25$ times \cite{tay2019pulsed}, datasets in this work are all recorded with $3$ averages. Data is acquired sequentially, each $y$-offset is held constant during which a sweep of all $x$-offsets is performed. 
Frequency and excitation amplitude stay constant until a dataset acquisition is completed.
Measurements are background corrected, \ac{TF} corrected for device independence and arranged by time samples and spatial offsets.
The two resulting raw data sets for each sequence type form the basis for the sequence generation process, yielding a low resolution system matrix ($\mat{\hat S}_\textup{LR}$ with \SI{0.67}{\mT} steps in $x$- and $y$-direction) and a high resolution system matrix ($\mat{\hat S}_\textup{HR}$ with \SI{0.5}{\mT} steps in $x$- and $y$-direction), respectively.

\begin{figure*}[!hb] 
    \centering
    \includegraphics[width=1.0\linewidth]{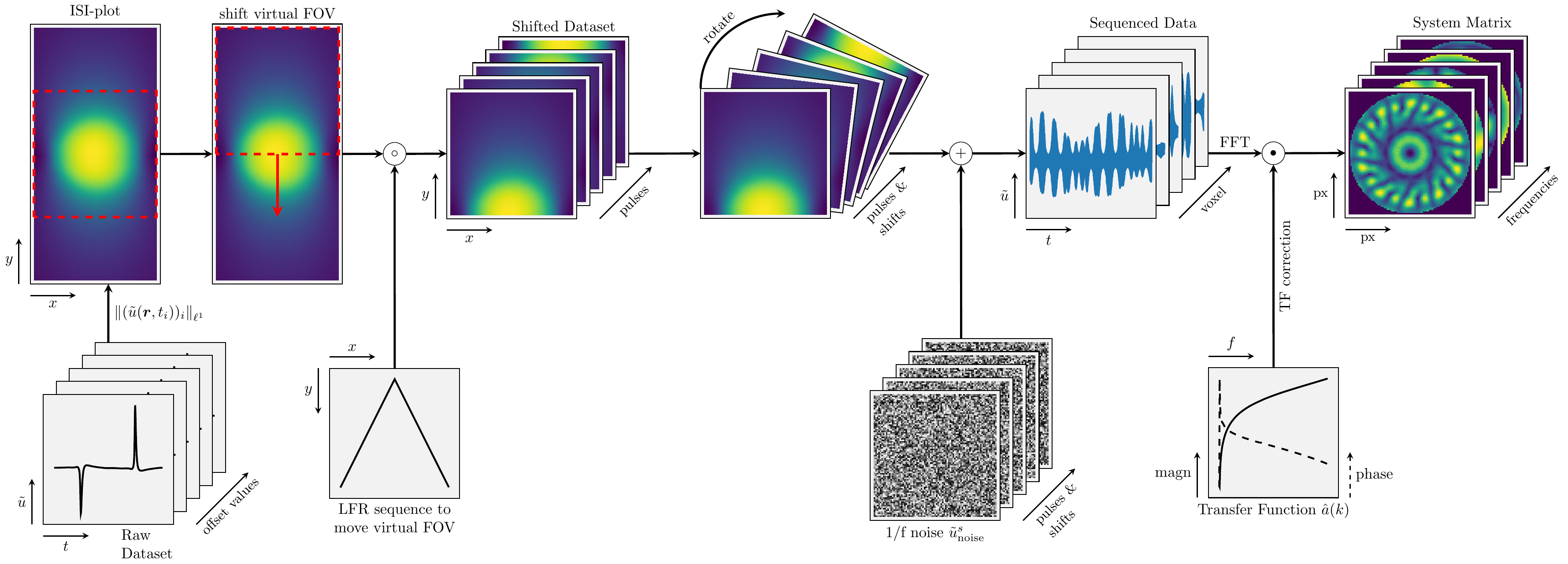}
    \caption{\textbf{Sequence generation diagram of the proposed shift-radial sequence.} Raw data is recorded with twice the offset values in $y$-direction than in $x$ (excitation direction) with pulsed excitation. The absolute values of $\tilde{u}(\mat{r},t)$ are summed up for each offset value to represent an \acf{ISI} plot. 
    To generate a sequence, a virtual \ac{FOV} is shifted from positive to negative $y$-offsets, mimicking an orthogonal shifting field. Simultaneously, it is continuously rotated to perform a rotation of the fields or the object. To avoid inverse crime, due to reusing the same dataset over and over, noise is added to mask repeating measurement background noise and prevent correlation from within the sequence. The sequenced data is fast Fourier transformed and \ac{TF} corrected to yield a system matrix, which is used to reconstruct images or create phantom spectra.
    Refer to Table \ref{tab:seq} for specific sequence parameter choices, excitation waveform and other measurement details.}
    \label{fig:seq_shiftradial}
\end{figure*}

\subsection{Sequence Generation}
\label{sec:sequence}

Three sequence types are presented in this paragraph, each based on two independent sets of measured raw data, to compose $\mat{\hat S}_\textup{LR}$ and $\mat{\hat S}_\textup{HR}$ for reconstruction. The parameters for excitation amplitude and waveform depend on the sequence version, as detailed in Table~\ref{tab:seq}.

The sequence itself is built from the time series data $\tilde{u}(\mat{r},t)$ in a way, that a subset of the dataset, like a virtual \ac{FOV}, is continuously shifted over the raw dataset by the \ac{LFR} sequence. If the sequence contains a rotation $\mat{R}(\theta(t))$, it is applied consecutively with the shifting $\mat{b}(t)$, as described by the rigid transformations
\begin{align*}
    \mathcal{T}(\mat{r},t) &= 
    \begin{cases}
        \mat{R}(\theta(t)) \, (\mat{r}+\mat{b}(t))  &\; \textup{seq. type }\, a)  \\
        \mat{R}(\theta(t)) \, \mat{r}               &\; \textup{seq. type }\, b)  \\
        \mat{r}+\mat{b}(t)                          &\; \textup{seq. type }\, c) 
    \end{cases} \\[1.5mm]
    \textup{for } \;
    \mat{R}(\theta(t)) &= 
    \begin{pmatrix}
        \cos \theta(t) & - \sin \theta(t) \\
        \sin \theta(t) &  \cos \theta(t)
    \end{pmatrix} , \;\,
    \mat{b}(t) =
    \begin{pmatrix}
        b_x (t) \\
        b_y (t)
    \end{pmatrix}  
\end{align*}

\noindent where $\theta(t)$, $b_x(t)$ and $b_y(t)$ are piecewise constant functions for each period.
The measurement noise within the transformed measurement voltage $\tilde{u}\!\left( \mathcal{T}(\mat{r},t),t \right)$ from the \ac{AWMPS}, would appear identically numerous times within the sequence, which can create reconstruction artifacts. 
To avoid correlation of measurement noise throughout the sequence, digitally generated noise is added with a higher noise level of

\begin{equation}
   \tilde{u}^s_\textup{noise}(t)  =  10 \, \sigma_\textup{meas} \, \tilde{u}_\textup{n,1/f}(t)
\end{equation}

\noindent which is crucial to mask the measurement noise floor of the raw dataset.
$\sigma_\textup{meas}$ is the standard deviation of the measurement background noise and $\tilde{u}_\textup{n,1/f}$ is the digitally generated noise, that has been fitted to the shape of the measured background noise. The overall shape resembles 1/f noise, which is typical for \acl{ICs} in the low frequency region.
The addition of the tenfold standard deviation is based on an analysis of the correlation coefficient versus the addition of \mbox{$n \cdot \sigma_\textup{meas}$} with $n \in \IR_+$, such that the noise in the generated sequence is dominated by uncorrelated noise.
In general, $\tilde{u}$ is the unprocessed voltage signal whereas the \ac{TF} corrected signal $u=a \ast \tilde{u}$ is in the domain of the magnetic moment.

Finally, the voltage signal is Fourier transformed (denoted by hat) and \ac{TF} corrected, yielding the system function 

\begin{equation}
    \hat s(\mat{r},k) = \frac{\hat a(k)}{T} \int_{0}^{T} \negthickspace \Big( \tilde{u}\!\left( \mathcal{T}(\mat{r},t),t \right) + \tilde{u}^s_{\textrm{noise}}(t) \Big) 
    e^\frac{-i2\pi kt}{T} dt 
\end{equation}

\noindent where $\mat{r} \in \IR^2$ is the position and $k \in \IN_0$ the frequency component. Negative frequency components are omitted, due to the symmetry of Fourier coefficients of real signals.
$T=J t_\textrm{max}$ is the sequence time length, $J$ the number of appended periods and $t_\textrm{max}$ the time length of a single period. Each voxel contains a time signal of $J$ periods.

\begin{figure*}[!t] 
    \centering
    \includegraphics[width=0.95\linewidth]{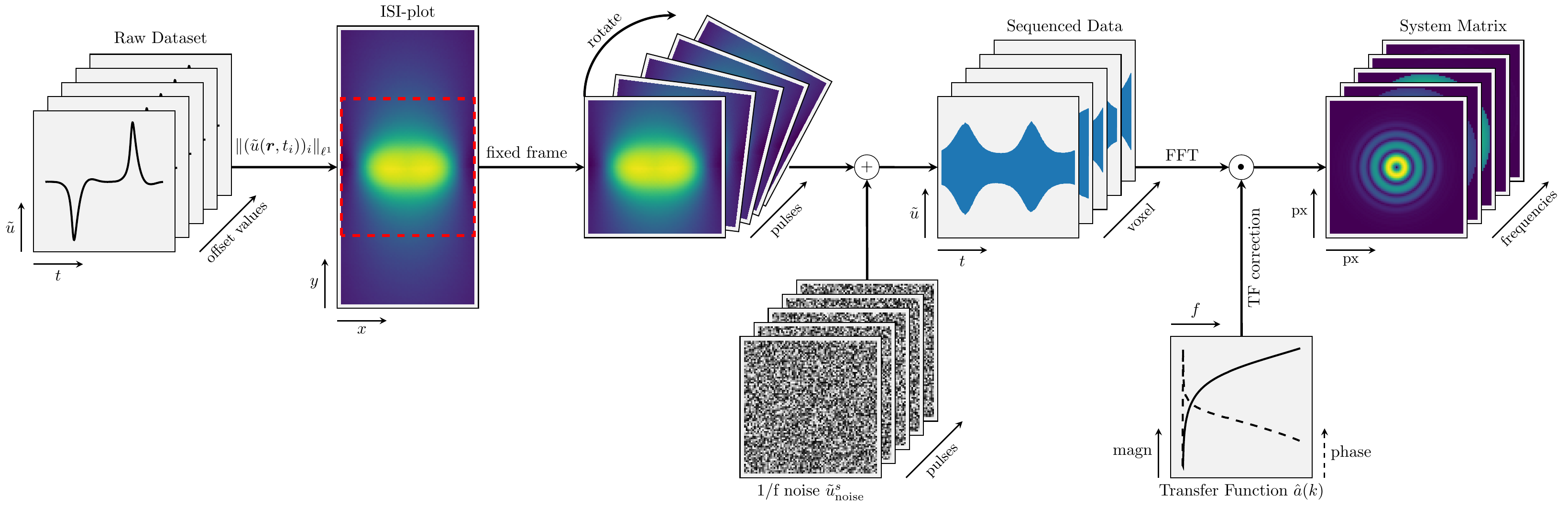}
    \caption{\textbf{Sequence generation diagram of the radial sequence.} Similar to the shift-radial sequence, the radial sequence is a simplified version, without the \ac{FOV} shifting component. A rectangular subset is solely rotated, as proposed by Knopp et al. in \cite{knopp2008trajectory}. In contrast to Fig. \ref{fig:seq_shiftradial} and Fig. \ref{fig:seq_meander}, the \ac{ISI}-plot shows the response to a sinusoidal excitation waveform (in $x$-direction). Otherwise, the system matrix is generated likewise, by noise addition, \ac{FFT} and \ac{TF} correction.
    Refer to Table \ref{tab:seq} for specific sequence parameter choices, excitation waveform and other measurement details.}
    \label{fig:seq_radialclassic}
\end{figure*}

\begin{figure*}[!b] 
    \centering
    \includegraphics[width=1.0\linewidth]{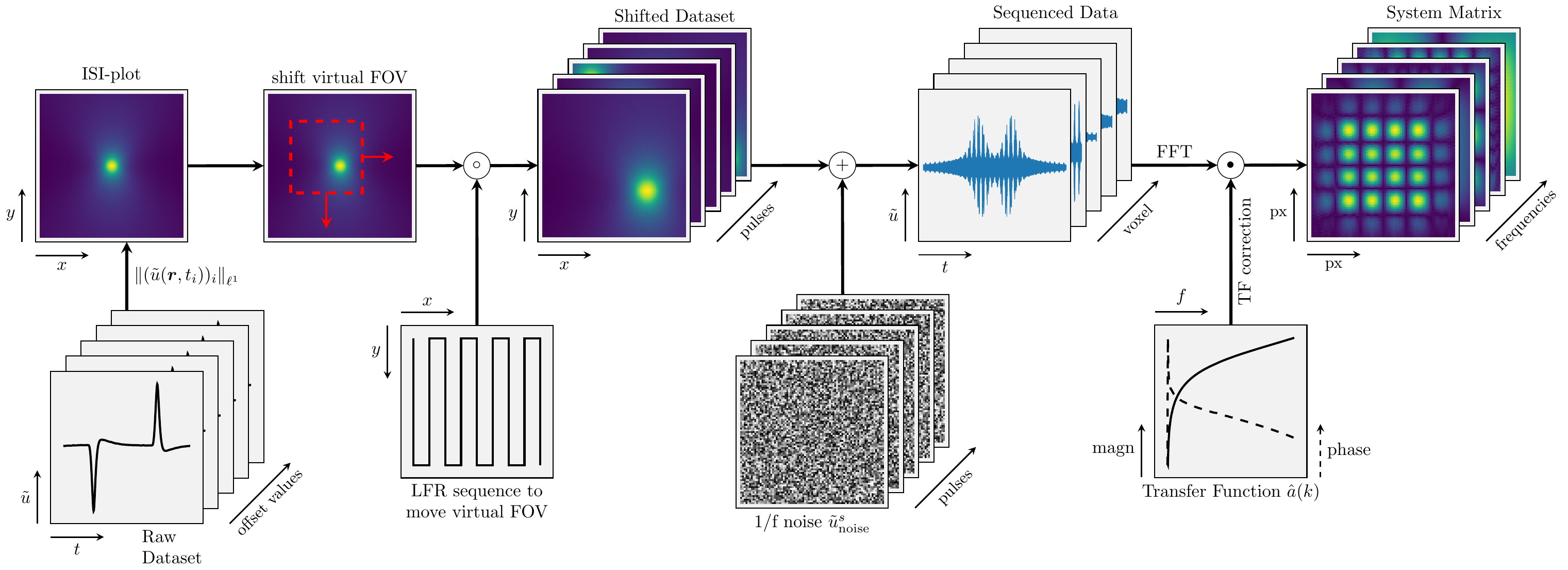}
    \caption{\textbf{Sequence generation diagram of the meander sequence proposed in \cite{tay2019pulsed}.}
    Measured raw data of 3 mT amplitude, with offsets in two spatial dimension is over-sampled in order to be able to move a virtual \ac{FOV}, which results in a smaller shifted dataset. The sequence name stems from the nature of this \ac{LFR} movement and the excitation is in $x$-direction only. The system matrix is generated identically as for all other sequences, by noise addition, \ac{FFT} and \ac{TF} correction.
    Refer to  Table \ref{tab:seq} for specific parameter choices of the \ac{LFR} sequence and other measurement details.}
    \label{fig:seq_meander}
\end{figure*}

To visualize the sequence generation process, a magnitude plot is used in Fig. \ref{fig:seq_shiftradial}, \ref{fig:seq_radialclassic} and  \ref{fig:seq_meander}. In this plot, the absolute value of the raw data $\tilde{u}(\mat{r},t)$ is summed up over a whole period for each offset value.
In other words, the signal intensity is integrated to quantify the system response for discrete, equally spaced measurement values $\tilde{u}_i = \tilde{u}(\mat{r},t_i)$ as in
\begin{equation}
    \norm{(\tilde{u}_i)_i}_{\ell^p} = \left( \sum_{i=0}^{i_\textrm{\scriptsize max}} \left| \tilde{u}_i\right|^p \right)^{1/p} .
\end{equation}

\noindent In this work, the $\ell ^1$-norm with $p=1$ is selected and referred to as the \acf{ISI}.
Thus, the \ac{ISI} is represented by bright yellow colors for positions of strong signal response and dark blue colors represent low signal response.
It is noted that the \ac{ISI}-plot is for visualization purposes only and the $\ell ^1$-norm is not implemented in the sequence calculation.
Within the virtual \ac{FOV}, the \ac{ISI}-plot visualizes a moving (and rotating) system response. The final sequence is generated by appending all time signals.
Noise addition, \acl{FFT} and \ac{TF} correction are done equivalently for all sequences.

\begin{figure}[H] 
    \centering
    \includegraphics[width=1.0\linewidth]{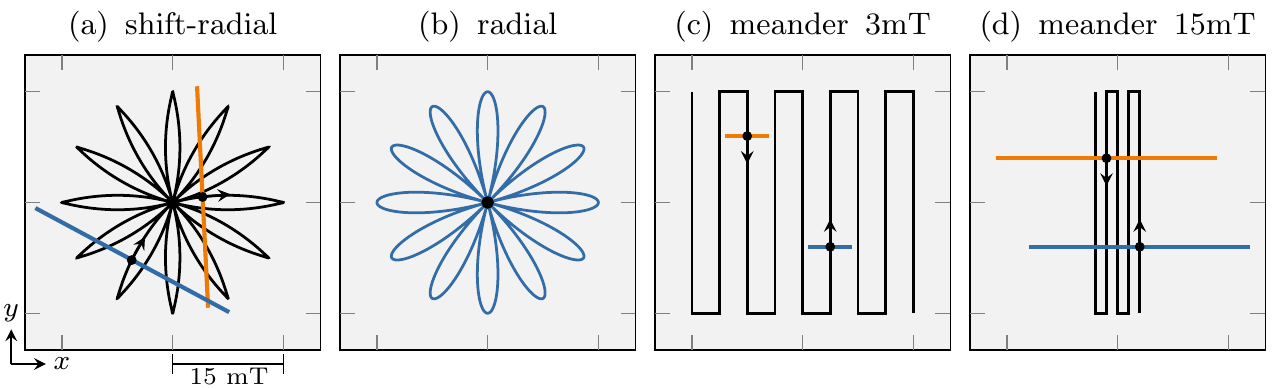}
    \caption{\textbf{Sequence trajectory plot.} The black line typifies the focus-field trajectory of the \ac{LFR}, as transformed by $\mathcal{T}(\mat{r},t)$. Orange and blue depict exemplary drive field excitations at individual time points, which then deflect the \ac{LFR} from this trajectory. The length of these bars correlates with the excitation amplitude and ticks are drawn in \SI{15}{\mT} increments. One side length equals \SI{40}{\mT}. Type (b) is not exposed to any offsets. Type (c) and (d) have identical \ac{FOV} boundaries. 
    Image sizes and scales are consistent throughout this work.
    For demonstration purposes, the density of the trajectories is reduced.}
    \label{fig:trajectory}
\end{figure}

\paragraph{Shift-radial sequence} Proposed in this work is a sequence with the properties of combining a shift, orthogonal to its excitation direction, with a rotation of the excitation itself, referred to as shift-radial sequence. The excitation direction in the raw data is in $x$-direction and the shift is in $y$-direction. The shift-radial sequence requires twice the oversampling of raw data along the $y$-direction with $\pm\SI{40}{\mT}$ than in $x$-direction with $\pm\SI{20}{\mT}$, to produce the necessary oversampling for the shift (see Fig. \ref{fig:seq_shiftradial}). 
A high excitation amplitude of $\SI{15}{\mT}$ is chosen, which corresponds to a wide \ac{PSF} spanning almost the entire \ac{FOV}. 
An \ac{LFR} sequence controls the virtual \ac{FOV} shift over the raw data to produce a shifted dataset with a given number of pulses per shift $J_\textup{PpS}$. These are rotated with a given number of shifts per rotation $J_\textup{SpR}$ with respect to their center. Subsequently, shifts and rotations are interweaved to generate a continuous acquisition of both in the time domain. 
In this work, $J_\textup{SpR}=32$ with $J_\textup{PpS}=31$ are chosen, amounting to $J=J_\textup{SpR} \cdot (J_\textup{PpS} \cdot 2-2) = 1920$ total periods. For each pulse the virtual \ac{FOV} is shifted one step in $y$-direction, starting from \SI{-35}{\mT} until $+$\SI{35}{\mT} is covered, where the direction inverses without sampling the turning-point nor the start-end-point twice (\SI{15}{\mT} amplitude of shifts, virtual FOV of \SI{40x40}{\mT}).
This sequence is similar to sequences known from MPI-FFL encoding and \ac{CT} and restores the lost information caused by large amplitudes \cite{Knopp2011d}. The trajectory is plotted in Fig. \ref{fig:trajectory} (a). 
For a better understanding of the influence of the excitation waveform, this sequence is generated for both, a sinusoidal and a pulsed rectangular excitation.

\paragraph{Radial sequence} The radial sequence is an adaptation of the shift-radial sequence, which solely utilizes a rotation of the raw data, without the virtual \ac{FOV} shift component. Hence, a square-shaped subset with $\pm\SI{20}{\mT}$ offsets in $x$- and $y$-direction is rotated with a given number of pulses per rotation $J_\textup{PpR}$ around its center, imitating a rotating excitation as proposed in Knopp et al. \cite{knopp2008trajectory}. The excitation amplitude is the same as the shift-radial sequence with $\SI{15}{\mT}$, but the waveform depicted in Fig. \ref{fig:seq_radialclassic} differs, as the radial sequence is only generated on the base of sinusoidal raw data. The sequence trajectory is shown in Fig \ref{fig:trajectory} (b).

\begin{table*}[!ht] 
\centering
\def\arraystretch{2.6}
\hspace*{0.0cm}
\caption{Overview of sequence details and measurement parameters of the AWMPS raw data.}
\label{tab:seq}
\resizebox{\textwidth}{!}{
\begin{tabular}{l|lll|l|ll|ll}
    & \multicolumn{3}{l}{drive field}  
    & 
    & \multicolumn{2}{l|}{sequence parameter}
    & \multicolumn{2}{l}{raw data matrix}
\\
    \pbox{2cm}{sequence\\name}
    & amplitude
    & waveform
    & frequency
    & ISI shape
    & sequence specific values
    & \pbox{2cm}{periods $J$\\ seq. time}
    & \ac{FOV} $x \times y$
    & \pbox{2cm}{grid low res.\\ grid high res.}
\\ \hline
meander 3mT     
    & \SI{3}{\mT}  
    & \pbox{2cm}{rectangular \\ \scriptsize{$t_\textup{r}=\SI{3}{\us}$}}
    & \SI{14.88}{\kHz} 
    & \pbox{2cm}{small,\\ localized}
    & \pbox{6.8cm}{$31$ steps in $x$ and $y$\\ \SI{1}{\mT} step resolution}
    & \pbox{2cm}{$1920$ \\ $\SI{129}{\ms}$ } 
    & \SI{80x80}{\mT}
    & \pbox{2cm}{$121 \times 121$\\ $161 \times 161$}
\\
meander 15mT       
    & \SI{15}{\mT} 
    & \pbox{2cm}{rectangular \\ \scriptsize{$t_\textup{r}=\SI{3}{\us}$}} 
    & \SI{14.88}{\kHz} 
    & \pbox{2cm}{broad area,\\ circular}
    & \pbox{6.8cm}{\vspace*{2.5mm} $31$ steps in $x$ and $y$\\ step res. in $x$: $\nicefrac{7}{31}$ \si{\mT}\\ step res. in $y$: \SI{1}{\mT}\vspace*{0.8mm}}
    & \pbox{2cm}{$1920$ \\ $\SI{129}{\ms}$ } 
    & \SI{80x80}{\mT}
    & \pbox{2cm}{$121 \times 121$\\ $161 \times 161$} 
\\
shift-radial  
    & \SI{15}{\mT}  
    & \pbox{2cm}{rectangular \\ \scriptsize{$t_\textup{r}=\SI{3}{\us}$}}
    & \SI{14.88}{\kHz} 
    & \pbox{2cm}{broad area,\\ circular}
    & \pbox{6.8cm}{\vspace*{2mm} 15 \si{\mT} amplitude of shifts\\$32$ shifts per rotation $J_\textup{SpR}$\\ $31$ pulses per shift $J_\textup{PpS}$\vspace*{0.8mm}} 
    & \pbox{2cm}{$1920$ \\ $\SI{129}{\ms}$ } 
    & \SI{40x80}{\mT}
    & \pbox{2cm}{$61 \times 121$\\ $81 \times 161$}
\\
shift-radial  
    & \SI{15}{\mT} 
    & sinusoidal 
    & \SI{14.88}{\kHz} 
    & \pbox{2cm}{broad area,\\ elliptic}
    & \pbox{6.8cm}{\vspace*{2mm} 15 \si{\mT} amplitude of shifts\\$32$ shifts per rotation $J_\textup{SpR}$\\ $31$ pulses per shift $J_\textup{PpS}$\vspace*{0.8mm}}
    & \pbox{2cm}{$1920$ \\ $\SI{129}{\ms}$ } 
    & \SI{40x80}{\mT}
    & \pbox{2cm}{$61 \times 121$\\ $81 \times 161$}
\\
radial  
    & \SI{15}{\mT} 
    & sinusoidal
    & \SI{14.88}{\kHz} 
    & \pbox{2cm}{broad area,\\ elliptic} 
    & \pbox{6.8cm}{$1920$ pulses per rotation $J_\textup{PpR}$}
    & \pbox{2cm}{$1920$ \\ $\SI{129}{\ms}$ } 
    & \SI{40x40}{\mT}
    & \pbox{2cm}{$61 \times 61$\\ $81 \times 81$}
    
\end{tabular}}
\end{table*}

\paragraph{Meander sequence} A Cartesian sequence is implemented with a discretization of $31$ steps in a resolution of $\SI{1}{\mT}$ per step in $x$- and $y$-direction, and is referred to as meander sequence due to its \ac{LFR} movement. By over-scanning the required square-shaped \ac{FOV} of $\pm \SI{20}{\mT}$, with a larger offset of $\pm\SI{40}{\mT}$, the initial measurement yields enough raw data to move the virtual \ac{FOV}, as illustrated in Fig. \ref{fig:seq_meander}. 
The virtual \ac{FOV} is moved along the meandering sequence, resulting in a shifted dataset.
The measurement data inside the shifted dataset has been moved along the trajectory and was sequentially appended to represent the particle response to the entire sequence. 
A version of the sequence is generated for each of the two drive field amplitudes, \SI{3}{\mT} and \SI{15}{\mT} ($x$-direction). The low excitation amplitude is chosen to be \SI{3}{\mT}, in a way to trade off \ac{SNR} and spatial resolution as proposed by Tay et. al \cite{tay2019pulsed}. 
In order to keep the \ac{FOV} boundaries identical between versions, the step resolution for \SI{15}{\mT} amplitude is reduced to $\nicefrac{7}{31}$ \si{\mT} in $x$, as illustrated in Fig. \ref{fig:trajectory} (d).
Both sequences consist of 31 steps in $x$- and 31 steps in $y$-direction, tracing the same path backwards without sampling the turning-point nor the start-end-point twice, amounting to $J=31 \cdot 31\cdot 2-2 = 1920$ periods. \\

The sequence types differ in the shape of their \ac{FOV}, depending on the fact if a rotation was applied. Rotating a rectangular dataset necessarily results in a circular sampled area (see Fig. \ref{fig:seq_shiftradial} and Fig. \ref{fig:seq_radialclassic}, far right), whereas for the meander-sequence (Fig. \ref{fig:seq_meander}) the \ac{FOV} stays rectangular. 
To create common ground between different sequence types, the same time length to compare sequence types is chosen, as this reflects the acquisition time. For a generated sequence length of \SI{129}{\ms} at \SI{14.88}{\kHz}, this amounts to $1920$ periods for each sequence in this work.

\subsection{Phantoms}
\label{sec:phantoms}

The intensity of the signal response of \ac{SPIONs} scales linearly with the concentration of the tracer \cite{Gleich2005Nature, Lu2013}. 
Consequently, the system matrix signal based on the \ac{AWMPS} needs to be scaled down.
One option is to introduce a scaling factor $\alpha_\textup{s}$ that scales the phantom concentration vector $\mat c_\textup{P} \in [0,1]^N$ ($N$ voxel), to a realistic value, in order to match the total iron mass within one voxel. $\alpha_\textup{s}$ together with a noise addition from a real life receiver system, results in realistic signal spectra. In all undiluted tracer experiments, a targeted tracer concentration of $c_\textup{target} =\SI{5}{\mg\of{Fe}\per\milli\liter}$ was used throughout this work. The concentration vector $\mat c$, which is used to create the phantom spectrum voltage $\mat{\hat u}_\textup{P}$ for reconstruction, is expressed by
\begin{align}
    \mat c &= \alpha_\textup{s}  \, \mat c_\textup{P} \nonumber \\
    &=  \frac{V_\textup{target}\; c_\textup{target} }{V_\textup{meas}\; c_\textup{meas}}  \, \mat c_\textup{P}
\end{align}
where $V_\textup{meas}$ is the volume and $c_\textup{meas}$ the concentration of the measured delta sample in the \ac{AWMPS}. An assumed linear gradient field $\mat G$, with $\det \mat G \neq 0$ in \si{\tesla\per\meter}, and a voxel grid vector $\mat g$ in \si{\mT} with 3 dimensions, yield the target concentration $V_\textup{target}$ in 

\begin{align}
    & V_\textup{target} = \prod_{i = 1}^{3} \left | \mat{G}^{-1} \, \mat g \right |_i \; = \SI{64}{\nano\liter} \; , \quad \textup{for}  \nonumber \\
    &
    \mat G = \begin{pmatrix}  \label{eq:G}
        -1.25 & 0 & 0 \\
        0 & -1.25 & 0 \\
        0 & 0 & 2.5
        \end{pmatrix} \;, \quad
    \mat g = \begin{pmatrix}
        0.5 \\
        0.5 \\ 
        1
       \end{pmatrix}
\end{align}

\noindent where $i$ denotes the $i$-th entry of the vector. This gradient field results in an exemplary scaling factor of $\alpha_\textup{s} = \num{9.4e-4}$, for a volume of \SI{20}{\micro \liter} and a concentration of \SI{17}{\milli\gram\of{Fe}\per\milli\liter} as described in Section \ref{sec:measurements}. Further experiments are performed with a lower 
\begin{figure}[H] 
    \centering
    \includegraphics[width=1.0\linewidth]{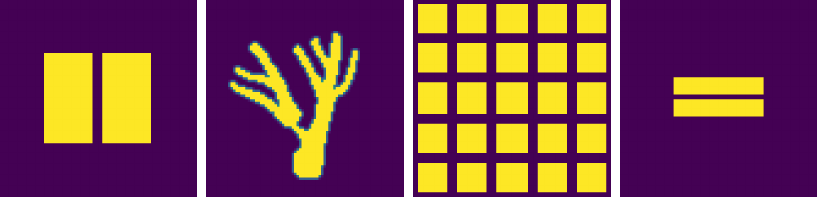}
    \caption{\textbf{Phantoms used in this study.} Reference for original shape. Left to right: Gap phantom with \SI{2.0}{\mT} (\SI{1.6}{\mm}) gap in $x$-direction, vessel phantom with stenosis in the left branch, large phantom to reveal borders of \ac{FOV} and fine resolution phantom with \SI{1.0}{\mT} (\SI{0.8}{\mm}) gap.}
    \label{fig:phantoms}
\end{figure}
\begin{figure*}[!b] 
    \centering
    \hspace*{-3mm}
    \includegraphics[width=0.95\textwidth]{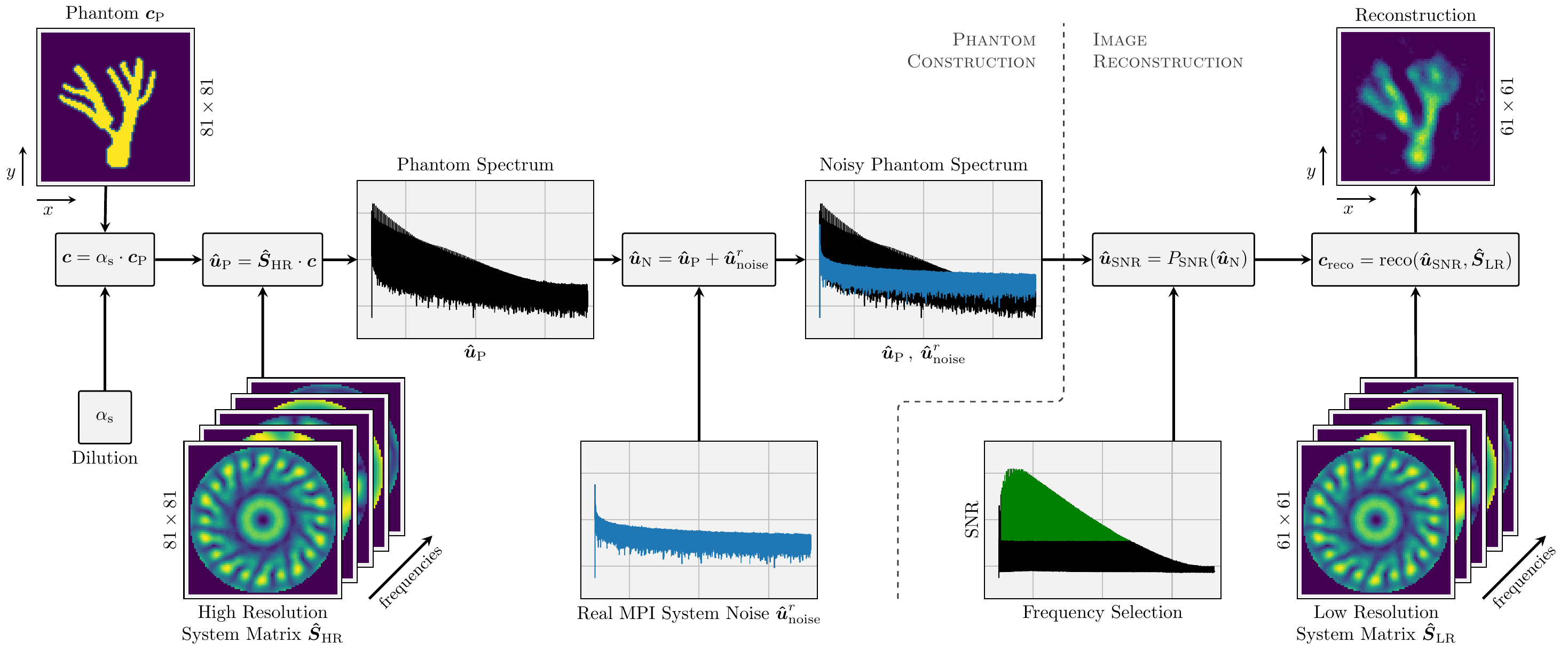}
    \caption{\textbf{Diagram of the implemented reconstruction process.} A vessel phantom with a visible stenosis is scaled to represent a tracer distribution $\mat{c}$ with a concentration of \SI{5}{\mg\of{Fe}\per\milli\liter} and a selectable dilution. The high resolution system matrix $\mat{\hat S}_\textup{HR}$ multiplied by the phantom vector $\mat{c}$ results in the phantom spectrum $\mat{\hat u}_\textup{P}$, which is overlaid by a realistic measurement noise level $\mat{\hat u}^r_\textup{noise}$. After frequency selection ($P_\textrm{SNR}$), the low resolution matrix $\mat{\hat S}_\textup{LR}$ is used to reconstruct the image using the iterative Kaczmarz method with Tikhonov regularization.
    All images span \SI{40x40}{\mT}.}
    \label{fig:reco}
\end{figure*}
\noindent phantom concentration $\mat c_\textup{P}$ to create a dilution series for a sensitivity analysis of all sequences types. Due to $\alpha_\textup{s}$, the change in $\mat{m}$ contributed by a voxel with concentration $\mat c_\textup{P}$ matches the real life magnetic moment change.

A second step involves a noise model to represent realistic MPI receiver noise during reconstruction. Real background noise measurements from a pre-clinical MPI scanner (Bruker, Ettlingen, Germany), utilizing an optimized \SI{42}{\milli \meter} coil \cite{Graeser2017SR}, are used to select a noise level of $n^r=\num{70e-{15}} \si{\A\m^2}$ for the highest frequency. Using a 1/f function to model the noise curve results in a higher noise for every frequency compared to the measured reference in \cite{GraeserHeadCoil2020}, guaranteeing a realistic \ac{SNR} scenario. The \ac{TF} corrected noise level added during image reconstruction is
\begin{align}
    \mat{\hat u}^r_\textup{noise} = \frac{n^r \zeta }{2\pi \; T_\textup{f,max}} \, \mat{T}  \; \in \IC^K
\end{align}
with $\zeta \sim \mathcal{C}\mathcal{N}(0,1)$ a standard complex normal random variable with zero mean and variance of $1$. $\mat{T} \in \IR_+^K$ contains the period time for each of the $K$ frequency components and $T_\textup{f,max}$ is the period length of the highest frequency in the receive band.
$\mat{\hat u}^r_\textup{noise}$ will overlay the scaled phantom spectrum $\mat{\hat u}_\textup{P}$, as visualized in Fig. \ref{fig:reco}, during the image reconstruction process. In Fig. \ref{fig:phantoms}, four phantoms are shown for original reference.

\subsection{Image Reconstruction}
\label{sec:reco}

Images are reconstructed using the iterative Kaczmarz method \cite{Kaczmarz1937}, which gives the calculated particle concentration $\mat{c} \in \IR^N_+$ for $N$ voxel, solving the linear system of equations
\begin{equation}
    \mat{\hat S} \:  \mat{c} = \mat{\hat u}_\textup{SNR}  
\end{equation}
where $\mat{\hat u}_\textup{SNR} \in \IC^K$ is a filtered subset of the emulated complex phantom voltage $\mat{\hat u}_\textup{P}$ with $K$ frequency components and $\mat{\hat S} \in \IC^{K \times N}$ is the system matrix in the frequency domain. 
For reconstruction the following least squares problem can be solved
\begin{equation} \label{eq:reco}
    \mat{c}^\lambda_\textup{reco} = \underset{\mat{c} \in \IR^N_+}{\textup{argmin}}\norm{   \mat{\hat S} \, \mat{c} - \mat{\hat u}_\textup{SNR} }^2_2 \, + \, \lambda \norm{\mat{c}}^2_2
\end{equation}
\begin{figure*}[!b] 
    \centering
    \includegraphics[width=1.0\linewidth]{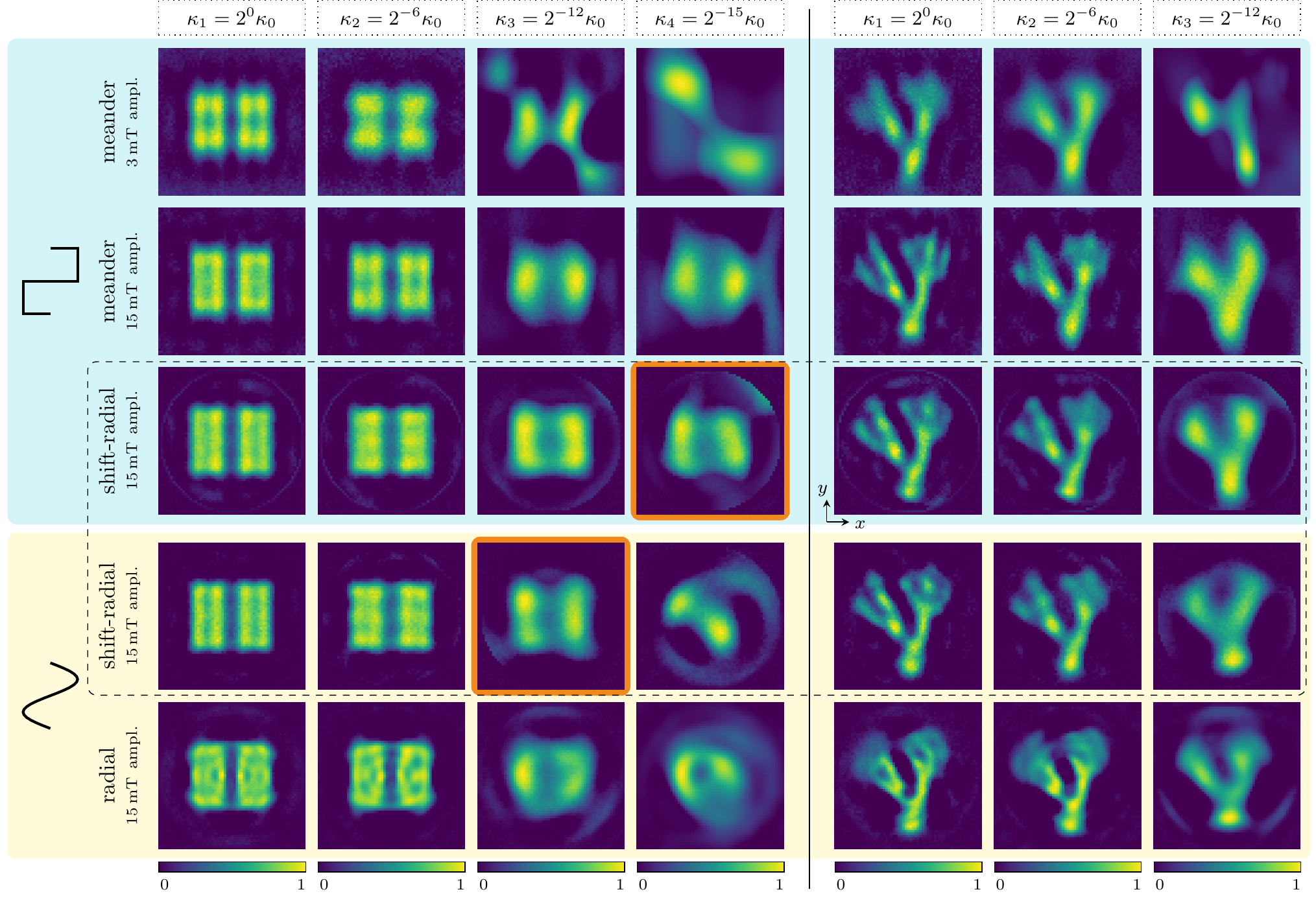}
    \caption{\textbf{Dilution series emulation for two excitation waveforms and three different sequence types.} Iron concentration $\kappa_0$ of \SI{5}{\mg\of{Fe} \per \milli\liter}, which is diluted from left to right, for two different phantoms. On the left, the phantom consists of two rectangles with a gap of \SI{2.0}{\mT} (\SI{1.6}{\mm}) in $x$-direction, whereas the right side depicts a vessel phantom with a stenosis towards the left branch (originals in Fig. \ref{fig:phantoms}). Reconstruction parameters are individually determined, to yield comparable noise in the image domain. System matrices are denoised and the intensity is normalized to 1 within each image. Highlighted in orange are reconstructions of similar quality, produced by the identical sequence but differing in excitation waveforms and dilution.
    All images span \SI{40x40}{\mT} (\SI{24.4x24.4}{\mm}), sequence trajectories are shown to scale in Fig. \ref{fig:trajectory}.}
    \label{fig:dilution}
\end{figure*}
where $\norm{ \mat{\hat S} \, \mat{c} - \mat{\hat u}_\textup{SNR} }^2_2$ is the data discrepancy term and $\norm{\mat c}^2_2$ is the penalization term used to dampen large oscillations in the solution. $\mat{c}^\lambda_\textup{reco}$ is the solution of the concentration distribution, controlled by a relative regularization parameter $\lambda \in \IR_+$, that blurs the image noise at the cost of spatial resolution.
With increasing dilution, the \ac{SNR} threshold is set higher and both, $\lambda$ and the number of iterations, increase. Additionally, the \ac{SNR} selection $P_\textup{SNR}$ is subjected to a minimal frequency threshold, acting as a high-pass, to filter out any signal before the first harmonic of the fundamental frequency. A reconstruction diagram is shown in Fig. \ref{fig:reco}, to visualize necessary steps. Handling noise amplification in images is described in detail in \cite{Boberg_2021}. 
We note that the solution of \eqref{eq:reco} in practice leads to better spatial resolutions than the width of the derivative of the particle magnetization curve would predict \cite{Knopp2011c}.

As this work contains emulated gradients only, distances in phantoms are best defined in mT. However, to give the reader a better classification of the results, the corresponding dimensions for a gradient as in \eqref{eq:G} are given in brackets e.g. \SI{2.0}{\mT} (\SI{1.6}{\mm}) in $x$- and $y$-direction. This same gradient is used for all emulations through-out this work. All phantoms are oriented in $xy$-plane.
Resulting images are reconstructed with the low resolution system matrix $\mat{\hat S}_\textup{LR}$, yielding an image resolution of 61 by 61 pixel, corresponding to a side length of \SI{40.0}{\mT} (\SI{24.4}{\mm}). 
Assuming the gradient field from \eqref{eq:G}, the fine system matrix $\mat{\hat S}_\textup{HR}$, that is used to constitute the phantom spectrum, accounts for an image resolution of $\SI{0.5}{\mT / pixel} \cdot (\SI{1.25}{\tesla \per \m})^{-1} = \SI{0.4}{\mm / pixel}$. System matrices are denoised to reduce the impact of system matrix noise on the reconstructed image \cite{weber2015enhancement}.

\begin{figure*}[!ht] 
    \centering
    \includegraphics[width=0.95\linewidth]{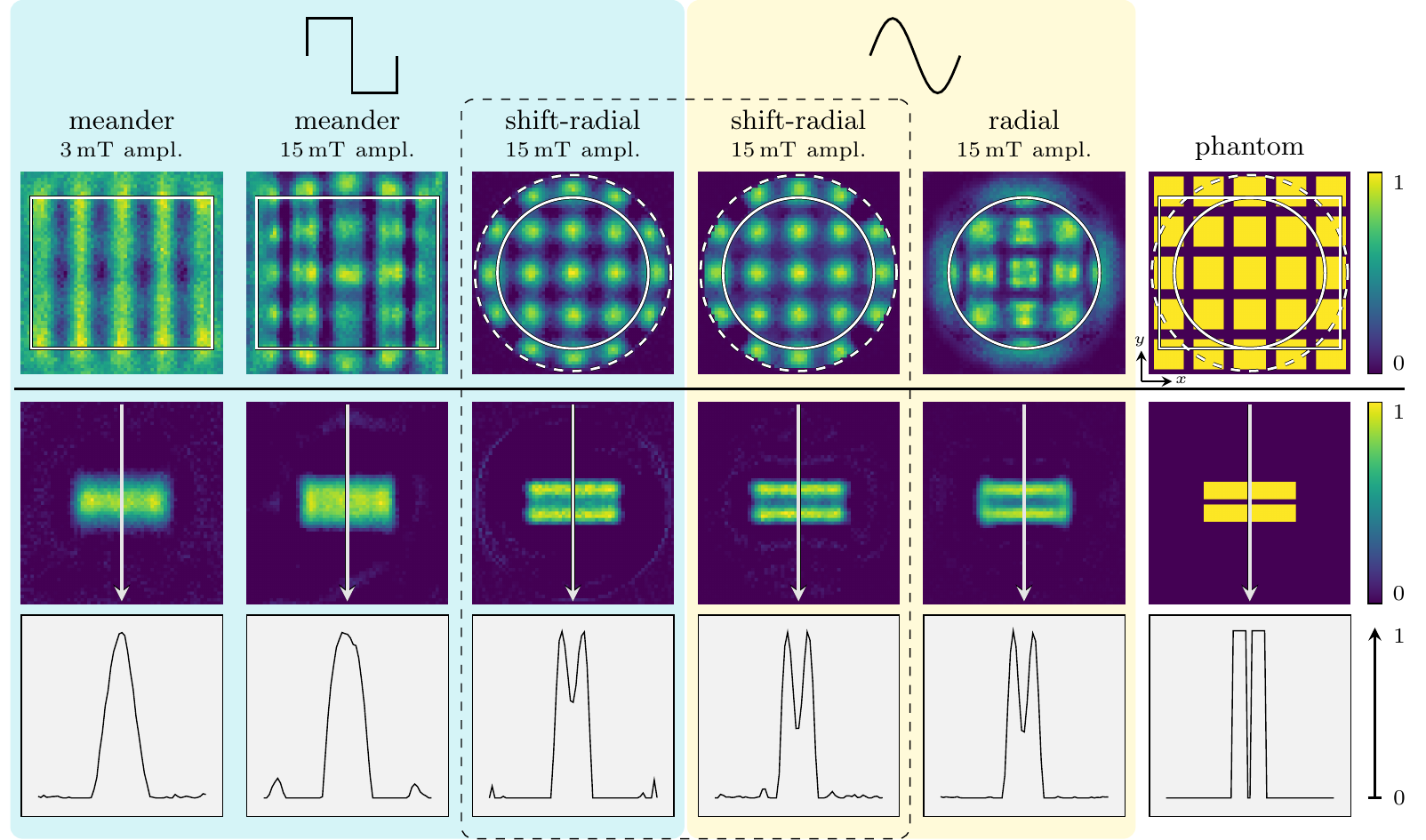}
    \caption{\textbf{Resolution simulation and \ac{FOV} comparison.} The top row compares the reconstruction of a 5 by 5 squares phantom, separated by \SI{2.0}{\mT} (\SI{1.6}{\mm}) gaps to distinguish the boundaries of the respective \ac{FOV}s. The phantom is shown in the rightmost column for reference. The shape of the \ac{FOV} depends on the sequence, resulting in a circular \ac{FOV} for radial sequence types. In the center row, a phantom with a \SI{1.0}{\mT} (\SI{0.8}{\mm}) gap in $y$-direction is shown, with the corresponding intensity profile in the bottom row. The iron concentration is \SI{5}{\mg\of{Fe} \per \milli\liter} each.}
    \label{fig:resolution}
\end{figure*}

\subsection{Implementation}
\label{sec:impl}

Raw data measurements in the \ac{AWMPS} are controlled by a system software implemented in the open source programming language \textit{Julia} \cite{bezanson2017julia}. It controls the parameter sweep through the chosen $x$-offset range and gradually steps up the $y$-offset for each excitation combination. The raw data is stored in an extended version of the Magnetic Particle Imaging data format (MDF) \cite{Knopp2016} and the sequence is afterwards calculated within a custom simulation framework developed in \textit{Julia}. System matrix reconstructions are based on the MPI reconstruction package developed in \cite{knopp2019mpireco}, accessible under \cite{MPIReco}.

\section{Results}
\label{sec:results}

Results are summarized in two main figures with phantom image reconstructions, one focusing on sensitivity by means of a dilution series (Fig. \ref{fig:dilution}), the other on resolution and \ac{FOV} shapes (Fig. \ref{fig:resolution}). 
Reconstruction parameters are individually determined to match the image noise across the different reconstructions.

Fig. \ref{fig:dilution} shows a dilution series of two phantoms to visually compare the results of different excitation waveforms, amplitudes and sequences, with an undiluted iron concentration of $\kappa_0 = \SI{5}{\mg\of{Fe} \per \milli\liter}$. The left part of the figure depicts a simple two bar phantom (gap in $x$-direction), the right part a more intricate vessel phantom with a stenosis in the left branch (see Fig. \ref{fig:phantoms} for originals).
For undiluted reconstructions, the shift-radial sequences and the meander sequence with high excitation amplitude perform very similar, whereas the low amplitude meander sequence suffers from a low \ac{SNR}. Fine structures are resolved independent of the excitation waveform. This effect becomes more visible with the intricacy of the vessel phantom on the right hand side in Fig. \ref{fig:dilution}. 
In undiluted reconstructions of the radial sequence, significant blurring appears towards the edges of the \ac{FOV} with the exception of a small region in the center, where resolution is excellent.
At a dilution of \SI{78}{\ug\per\milli\liter} (1:64, $\kappa_2$), the pulsed \SI{3}{\mT} meander sequence starts to lose the ability to resolve any fine structures and for dilutions of \SI{1.2}{\ug\per\milli\liter} (1:4096, $\kappa_3$) it stops to resemble even the rough shape of the original phantom. However, the pulsed \SI{15}{\mT} meander sequence is able to display basic phantom features down to the highest dilution.
The shift-radial sequence is able to depict phantom outlines of simple shapes at $\kappa_3$ for both, pulsed and sinusoidal excitation. The pulsed excitation for extremely high dilutions of \SI{153}{\ng \per \milli\liter} (1:32768, $\kappa_4$) seems to have a slight advantage over the sinusoidal excitation, being still able to marginally reconstruct a simple phantom whereas the sinusoidal excitation is entirely dominated by system noise. When comparing pulsed sequences with \SI{15}{\mT} amplitude, edges are traced more accurately by the pulsed shift-radial sequence than the meander sequence at dilution $\kappa_3$ for the left phantom (row 2 and 3).
The sinusoidal radial sequence is able to resolve most structures at $\kappa_2$, especially around its center, however it becomes heavily distorted beyond $\kappa_3$.

In Fig. \ref{fig:resolution}, the characteristic \ac{FOV}s of all sequences are visualized using a phantom that extends over the full image width ($\SI{40}{\mT}$). For the meander sequence, the \ac{FOV} spans to the out-most \ac{LFR} boundaries ($\SI{36x30}{\mT}$). Shift-radial is confined with full resolution to the radius of its shift amplitude (inner 
\begin{figure}[H] 
    \centering
    \includegraphics[width=0.85\linewidth]{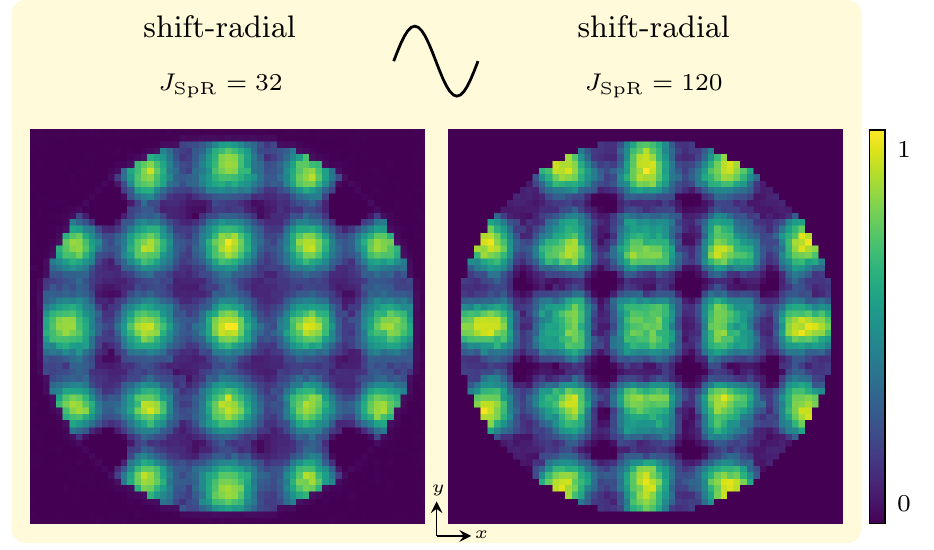}
    \caption{\textbf{Trajectory density comparison.} The image to the left stems from the same system matrix for the sinusoidal shift-radial sequence throughout this work, with the sequence parameter of $J_\textup{SpR}$ = 32 shifts per rotation. The image to the right has a denser trajectory with $J_\textup{SpR}$ = 120, consequently the sequence is longer with 7200 periods instead of 1920. Image reconstruction parameters are identical.}
    \label{fig:spr120}
\end{figure}
\noindent circle, \SI{15}{\mT}), although this extends with $\sqrt{2}$ to a second region of slightly degraded reconstruction ability (dashed, white line). 
The shift in excitation, which can be described as a pulse line, results in a rectangular shaped sampling area. As this rectangle is rotated, only a circle with the diameter of one side length is fully covered. As shown in Fig. \ref{fig:resolution}, the area in between the inner circle with diameter of a side length and a circle with a diameter of a diagonal length, is only occasionally covered by the corner of the rectangular sampling area. Thus, the resolution decreases as sampling becomes sparse. 
The radial sequence shows a very fine resolution in the center of the \ac{FOV}, which deteriorates continuously with increasing radius. As the excitation is rotated without shifting, the center of the \ac{FOV} is sampled by every pulse, which in turn results in a fine local resolution. This comes with the drawback of losing precision towards the edges of the \ac{FOV}. 
A bar-phantom with a \SI{1.0}{\mT} (\SI{0.8}{\mm}) gap in $y$-direction is shown in the center row of Fig. \ref{fig:resolution}, with the associated intensity profile over $y$ in the bottom row. An average of three pixel left and right of the center line is taken to calculate the intensity plot. 
The reconstruction based on the meander sequence is not able to resolve the two bars in both cases, whereas shift-radial and radial sequences are able to do so. This can be explained by the limitation of Cartesian sequences to be an-isotropic in their nature, as the excitation is oriented solely in one direction. In this work, $x$ is chosen for excitation and therefore a better resolution in $x$ is obtained than in $y$, as also apparent in the an-isotropic \ac{FOV} plot in the top row for the meander sequence, independent of using small or large amplitude.

In Fig. \ref{fig:spr120} the impact of changing a sequence parameter is shown by comparing shift-radial sequences with different trajectory densities. On the left, the identical system matrix that was used throughout this work for sinusoidal shift-radial reconstructions with $J_\textup{SpR}= 32$ shifts per rotation was implemented. On
\begin{figure}[H] 
    \centering
    \includegraphics[width=1.0\linewidth]{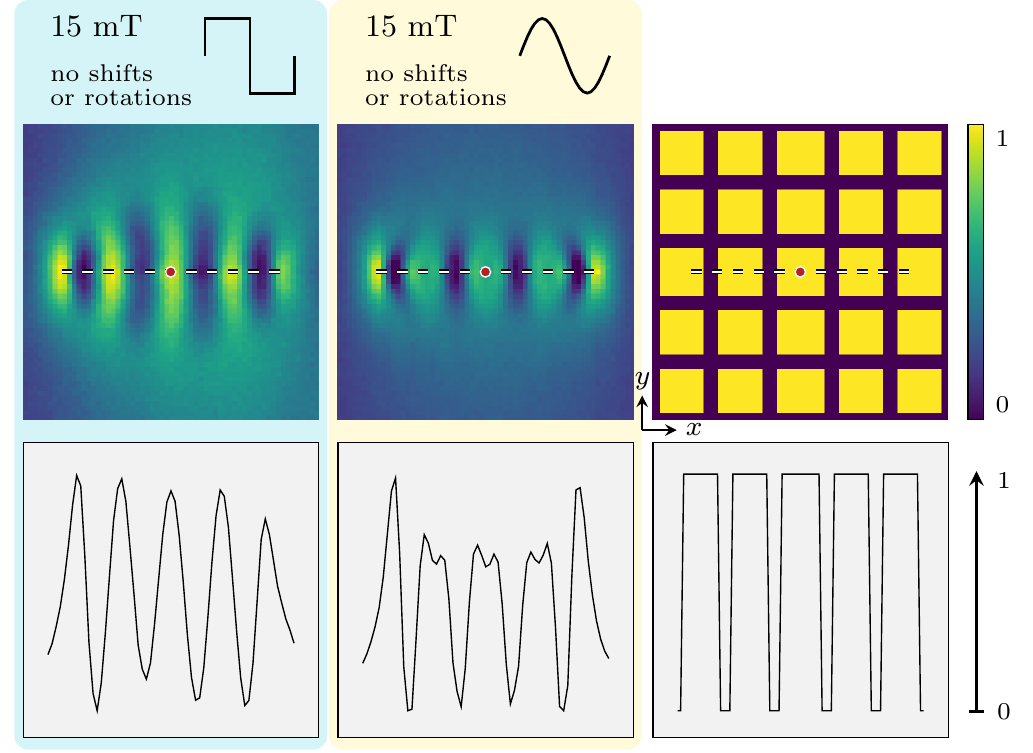}
    \caption{\textbf{Comparison of pulsed and sinusoidal data without a sequence trajectory.} 
    In order to reveal the effect of information encoded within the phase of the measured raw data for different excitation waveforms, an immobile sequence is generated that does not use shifts or rotations of any kind (also $1920$ periods). Consequently, the \ac{LFR} is only moved by the drive field (in $x$) and the resulting \ac{FOV} is a single line with \SI{15}{\mT} amplitude (white, dashed). The images span \SI{40x40}{\mT}.
    Below, the intensity profile along this line in $x$ is plotted.}
    \label{fig:noTraj}
\end{figure}
\noindent the right side, a sequence was generated with $J_\textup{SpR} = 120$ that has $7200$ periods instead of $1920$. All other parameters and reconstructions are identical. Results indicate that for a good representation of edges and corners a high density is advantageous. 

In a final result in Fig. \ref{fig:noTraj}, the effect that phase information from within our measured data has on resolution, is isolated by creating an immobile sequence without any shifted or rotated components in the trajectory. This can be done by setting the shift amplitude and step resolution to zero and the  repetitions to $1920$ periods, so the time length is identical to all other sequences. 
The lack of any focus-field movement implies that the \ac{LFR} is solely moved by the excitation waveform along a single line in $x$, resulting in a \SI{30}{\mT} \ac{FOV} line (dashed, white line).
Shown in Fig. \ref{fig:noTraj} are reconstructions based on pulsed and sinusoidal data, which are both capable of restoring a simple pattern in the \ac{FOV}. The elliptic outline of this area is shaped similar to the outline of the \ac{ISI} plots of raw data in Fig. \ref{fig:seq_shiftradial} and \ref{fig:seq_radialclassic} for pulsed and sinusoidal data respectively.

\section{Discussion}
\label{sec:discussion}

By combining a shifting focus field with a rotating drive field of large amplitude, the proposed shift-radial sequence overcomes previous limitations in amplitude and acquisition time for pulsed rectangular excitations.
Shift-radial sequences provide an alternative to Cartesian sampling schemes, due to their ability to resolve \SI{1.0}{\mT} gaps (\SI{0.8}{\mm} for a gradient of ($-1.25$, $-1.25$, $2.5$) \si{\tesla \per \m}) and their spatially isotropic sampling trajectory.

Furthermore, pulsed excitation seems to bear a slight advantage in sensitivity compared to sinusoidal excitation for the identical shift-radial sequence, if the two images marked in orange in Fig. \ref{fig:dilution} are considered reconstructions of the same quality. Note that a factor of $10^3$ distinguishes column 3 from 4. The basis of this advantage lies in the simultaneous response of many spatial positions, resulting in a better \ac{SNR}. The magnitude of this advantage becomes visible at higher dilutions only, and it depends on the phantom or on the medical application. The more homogeneous the distribution of \ac{SPIONs} (e.g. perfusion imaging), the higher seems the advantage of pulsed excitation in contrast to the accumulation of tracer material in a bulk (e.g. stem cell tracking).

The drive field frequency was chosen to be \SI{14.88}{kHz} as this is the upper limitation of the amplifier to produce a clean waveform with rise-times around \SI{3}{\us}. The relaxation behaviour of the used tracer (Perimag,  Fig. \ref{fig:plot_freq}) allows the use of \SI{14.88}{kHz} in this work, due to its quick relaxation.
Two points need to be clarified regarding this choice:
First, tracers need to be asserted individually and measurement parameters such as frequency and rise time would need to be adapted for larger particles to match their relaxation. Tay et al. proposed the pulsed excitation scheme for large particles beyond a specific size of about \SI{28}{\nano\meter} \cite{tay2019pulsed,Shasha2019NanoparticleCS} to circumvent limiting relaxation effects and improve resolution. The so called ``relaxation wall'' blurs the \ac{MPI} image, as particles do not follow the Langevin model \cite{Tay_relaxation_wall_2017}, and the optimum predicted resolution cannot be achieved \cite{Ferguson2009_coresize, weizenecker_mirco-magnetic_2012}. One precondition in order to reach optimum resolution for any particle system is to ensure that the particle magnetization is sufficiently saturated to generate a strong signal, which sets an upper limit to the excitation frequency. 
The second point is that due to the quick relaxation of Perimag and the sigmoid-shaped excitation, which differs from an idealized rectangle, a small time-shift is observable in the raw data, as shown in Fig. \ref{fig:plot_freq} on the left. Therefore, the phase information of the signal is distinguishable within the measured raw data, despite our assumption that all phase information is lost by pulsed excitation. Due to this fact, the following argument needs to be considered: Results in Fig. \ref{fig:dilution}, \ref{fig:resolution} and \ref{fig:spr120} show the combined effect of pulsed excitation including phase and the effect of the sequence trajectory for system matrix reconstruction. A separation of individual contributions is not possible with our measurement data.
Ultimately, system matrix reconstruction of pulsed sequences with high amplitude combine the benefit in the \ac{SNR} and profit from any phase information within the particle response. This holds true for meander as well as shift-radial sequences and explains why a meander sequence with \SI{15}{\mT} amplitude performs similar compared to the pulsed shift-radial sequence (Fig. \ref{fig:dilution}, rows 2 and 3) for our measurement data and using system matrix reconstruction.
Shift-radial sequences profit from phase information, but they would be capable to restore basic phantom features even without any phase information, similar to a Radon transform \cite{Knopp2011d}.
For an ideal rectangular excitation (no phase discrimination) Cartesian meander sequences depend on a small excitation amplitude for achieving high resolution.
For this reason a low amplitude (\SI{1}{\mT}) was chosen in previous works \cite{tay2019pulsed} to guarantee a small sampling kernel and sustain high resolution, with the draw back of a low \ac{SNR}.

The sole benefit due to phase information for pulsed and sinusoidal raw data can be identified in Fig. \ref{fig:noTraj}.
Surprisingly, basic phantom features are resolved by the reconstruction based on pulsed raw data, which can only be attributed to phase information as there is no additional information provided by means of a sequence trajectory. Edges of the squares along the white \ac{FOV} line are slightly better shaped for sinusoidal data, when the intensity profiles along the $x$-center line are compared. However, the pulsed reconstruction clearly separates all $5$ squares, relying only on the small time-shift that is induced in the raw signal, as shown in Fig. \ref{fig:plot_freq} on the left. If this holds true for larger particle systems remains to be proven. As a consequence, system matrix reconstruction may be able to sustain basic resolution even for large single-core particles, if the Nyquist criterion is not violated during waveform slopes (rise-time).
Resolution improvements of pulsed excitation compared to sinusoidal excitation were neither observed nor expected for the given tracer and measurement parameters. However, the proposed shift-radial imaging sequence and reconstruction method does not make any assumptions on the particle size and thus we predict that it generalizes to different relaxations and larger particles, if individual particle dynamics are heeded and an equilibrium state is reached.

Concerning our implemented methods, the noise added during sequence generation, as well as the noise added during image reconstruction, are overestimated and higher than in a well designed scanner system.
While rectangular excitation can be superior for large particles, the results in this work could not assert a clear advantage for standard MPI tracers with a size lower than the relaxation wall. In fact, sinusoidal excitation seems superior at high concentrations, as it is less prone to artifacts at the edges of the FOV.
The signal intensity for each spatial position is linked to the derivative of the excitation field at the time the \ac{LFR} passes \cite{Goodwill2010}. For sinusoidal excitation this leads to an attenuation of the signal at the \ac{FOV} border, which is lower for pulsed sequences. This attenuation may help avoiding artifacts at the hard cut off at the end of the circular \ac{FOV} for the presented radial-shift sequences.
Sinusoidal drive fields performed equally in almost all regards throughout this study, hence the effort for a pulsed non-resonant scanner may only be reasonable for larger particle systems. Further research on the application of the proposed method to large single-core particles has to be done.
Improvements may be limited by \acl{PNS} and the \acl{SAR}, reducing the advantage of high drive field amplitudes for in-vivo imaging in large animals or humans. 
Also, the optimum frequency-amplitude combination for shift-radial and the choice of sequence parameters to trade off resolution and acquisition speed need to be investigated.

Although rectangular excitation patterns did not outperform the sinusoidal ones for medium-sized MPI particles in this work, they certainly have the potential to improve MPI not only for large particles. By allowing to vary the excitation frequency as an adjustable hardware parameter, one can tailor the excitation pattern to the currently used particle system, giving maximum flexibility. Potentially, this allows optimizing the contrast in multi-contrast MPI applications, e.g. for discriminating different particle sizes \cite{Shasha2019}.


\bibliographystyle{IEEEtran}
\bibliography{MPILiterature.bib}

\begin{thebibliography}{10}
\providecommand{\url}[1]{#1}
\csname url@samestyle\endcsname
\providecommand{\newblock}{\relax}
\providecommand{\bibinfo}[2]{#2}
\providecommand{\BIBentrySTDinterwordspacing}{\spaceskip=0pt\relax}
\providecommand{\BIBentryALTinterwordstretchfactor}{4}
\providecommand{\BIBentryALTinterwordspacing}{\spaceskip=\fontdimen2\font plus
\BIBentryALTinterwordstretchfactor\fontdimen3\font minus
  \fontdimen4\font\relax}
\providecommand{\BIBforeignlanguage}[2]{{%
\expandafter\ifx\csname l@#1\endcsname\relax
\typeout{** WARNING: IEEEtran.bst: No hyphenation pattern has been}%
\typeout{** loaded for the language `#1'. Using the pattern for}%
\typeout{** the default language instead.}%
\else
\language=\csname l@#1\endcsname
\fi
#2}}
\providecommand{\BIBdecl}{\relax}
\BIBdecl

\bibitem{Gleich2005Nature}
B.~Gleich and J.~Weizenecker, ``Tomographic imaging using the nonlinear
  response of magnetic particles,'' \emph{Nature}, vol. 435, no. 7046, pp.
  1214--1217, 2005.

\bibitem{Panagiotopoulos2015}
N.~Panagiotopoulos, R.~L. Duschka, M.~Ahlborg, G.~Bringout, C.~Debbeler,
  M.~Graeser, C.~Kaethner, K.~L{\"{u}}dtke-Buzug, H.~Medimagh, J.~Stelzner,
  T.~M. Buzug, J.~Barkhausen, F.~M. Vogt, and J.~Haegele, ``Magnetic particle
  imaging: Current developments and future directions,'' \emph{International
  Journal of Nanomedicine}, vol.~10, pp. 3097--3114, 2015.

\bibitem{Knopp2010PhysMedBio}
T.~Knopp, J.~Rahmer, T.~F. Sattel, S.~Biederer, J.~Weizenecker, B.~Gleich,
  J.~Borgert, and T.~M. Buzug, ``Weighted iterative reconstruction for magnetic
  particle imaging,'' \emph{Physics in Medicine and Biology}, vol.~55, no.~6,
  pp. 1577--1589, 2010.

\bibitem{Graeser2013}
M.~Graeser, T.~Knopp, M.~Gr{\"{u}}ttner, T.~F. Sattel, and T.~M. Buzug,
  ``Analog receive signal processing for magnetic particle imaging,''
  \emph{Medical Physics}, vol.~40, no.~4, p. 42303, 2013.

\bibitem{Vogel2019Micro}
P.~Vogel, M.~A. R{\"{u}}ckert, S.~J. Kemp, A.~P. Khandhar, R.~M. Ferguson,
  S.~Herz, A.~Vilter, P.~Klauer, T.~A. Bley, K.~M. Krishnan, and V.~C. Behr,
  ``Micro-traveling wave magnetic particle imaging - sub-millimeter resolution
  with optimized tracer ls-008,'' \emph{IEEE Transactions on Magnetics},
  vol.~55, no.~10, pp. 1--7, 2019.

\bibitem{Weizenecker2009}
J.~Weizenecker, B.~Gleich, J.~Rahmer, H.~Dahnke, and J.~Borgert,
  ``Three-dimensional real-time in vivo magnetic particle imaging,''
  \emph{Physics in Medicine and Biology}, vol.~54, no.~5, pp. L1--L10, 2009.

\bibitem{Graeser_2020}
\BIBentryALTinterwordspacing
M.~Graeser, P.~Ludewig, P.~Szwargulski, F.~Foerger, T.~Liebing, N.~D. Forkert,
  F.~Thieben, T.~Magnus, and T.~Knopp, ``Design of a head coil for high
  resolution mouse brain perfusion imaging using magnetic particle imaging,''
  \emph{Physics in Medicine {\&} Biology}, vol.~65, no.~23, p. 235007, dec
  2020. [Online]. Available: \url{https://doi.org/10.1088/1361-6560/abc09e}
\BIBentrySTDinterwordspacing

\bibitem{graeser2019Head}
M.~Graeser, F.~Thieben, P.~Szwargulski, F.~Werner, N.~Gdaniec, M.~Boberg,
  F.~Griese, M.~M{\"{o}}ddel, P.~Ludewig, D.~van~de Ven, O.~M. Weber,
  O.~Woywode, B.~Gleich, and T.~Knopp, ``Human-sized magnetic particle imaging
  for brain applications,'' \emph{Nature Communications}, vol.~10, no.~1, 2019.

\bibitem{Mason2017}
E.~E. Mason, C.~Z. Cooley, S.~F. Cauley, M.~A. Griswold, S.~M. Conolly, and
  L.~L. Wald, ``Design analysis of an mpi human functional brain scanner.''
  \emph{International journal on magnetic particle imaging}, vol.~3, no.~1,
  2017.

\bibitem{Rahmer2018}
J.~Rahmer, C.~Stehning, and B.~Gleich, ``Remote magnetic actuation using a
  clinical scale system,'' \emph{PLoS ONE}, vol.~13, no.~3, p. e0193546, 2018.

\bibitem{Haegele2013b}
J.~Haegele, S.~Biederer, H.~Wojtczyk, M.~Gr{\"{a}}ser, T.~Knopp, T.~M. Buzug,
  J.~Barkhausen, and F.~M. Vogt, ``Toward cardiovascular interventions guided
  by magnetic particle imaging: First instrument characterization,''
  \emph{Magnetic Resonance in Medicine}, vol.~69, no.~6, pp. 1761--1767, 2013.

\bibitem{Wegner2019}
F.~Wegner, T.~Friedrich, A.~von Gladiss, U.~Grzyska, M.~M. Sieren,
  K.~Lüdtke-Buzug, A.~Frydrychowicz, T.~M. Buzug, J.~Barkhausen, and
  J.~Haegele, ``Magnetic particle imaging: Artifact-free metallic stent lumen
  imaging in a phantom study,'' \emph{{CardioVascular} and Interventional
  Radiology}, vol.~43, no.~2, pp. 331--338, oct 2019.

\bibitem{Szwargulski2020ACS}
\BIBentryALTinterwordspacing
P.~Szwargulski, M.~Wilmes, E.~Javidi, F.~Thieben, M.~Graeser, M.~Koch,
  C.~Gruettner, G.~Adam, C.~Gerloff, T.~Magnus, T.~Knopp, and P.~Ludewig,
  ``Monitoring intracranial cerebral hemorrhage using multicontrast real-time
  magnetic particle imaging,'' \emph{ACS Nano}, vol.~14, no.~10, pp.
  13\,913--13\,923, 2020, pMID: 32941000. [Online]. Available:
  \url{https://doi.org/10.1021/acsnano.0c06326}
\BIBentrySTDinterwordspacing

\bibitem{GraeserHeadCoil2020}
\BIBentryALTinterwordspacing
M.~Graeser, P.~Ludewig, P.~Szwargulski, F.~Foerger, T.~Liebing, N.~D. Forkert,
  F.~Thieben, T.~Magnus, and T.~Knopp, ``Design of a head coil for high
  resolution mouse brain perfusion imaging using magnetic particle imaging,''
  \emph{Physics in Medicine {\&} Biology}, vol.~65, no.~23, p. 235007, 2020.
  [Online]. Available: \url{http://dx.doi.org/10.1088/1361-6560/abc09e}
\BIBentrySTDinterwordspacing

\bibitem{Ludewig2017Stroke}
P.~Ludewig, N.~Gdaniec, J.~Sedlacik, N.~D. Forkert, P.~Szwargulski, M.~Graeser,
  G.~Adam, M.~G. Kaul, K.~M. Krishnan, R.~M. Ferguson, A.~P. Khandhar,
  P.~Walczak, J.~Fiehler, G.~Thomalla, C.~Gerloff, T.~Knopp, and T.~Magnus,
  ``Magnetic particle imaging for real-time perfusion imaging in acute
  stroke,'' \emph{ACS Nano}, vol.~11, no.~10, pp. 10\,480--10\,488, 2017.

\bibitem{rahmer2015first}
J.~Rahmer, A.~Halkola, B.~Gleich, I.~Schmale, and J.~Borgert, ``First
  experimental evidence of the feasibility of multi-color magnetic particle
  imaging,'' \emph{Physics in Medicine \& Biology}, vol.~60, no.~5, p. 1775,
  2015.

\bibitem{Shasha2019}
C.~Shasha, E.~Teeman, K.~M. Krishnan, P.~Szwargulski, T.~Knopp, and
  M.~M{\"{o}}ddel, ``Discriminating nanoparticle core size using multi-contrast
  mpi,'' \emph{Physics in Medicine and Biology}, vol.~64, no.~7, p. 74001,
  2019.

\bibitem{stehning2016simultaneous}
C.~Stehning, B.~Gleich, and J.~Rahmer, ``Simultaneous magnetic particle imaging
  (mpi) and temperature mapping using multi-color mpi,'' \emph{International
  Journal on Magnetic Particle Imaging}, vol.~2, no.~6, pp. 1--6, 2016.

\bibitem{M_ddel_2018}
M.~M{\"{o}}ddel, C.~Meins, J.~Dieckhoff, and T.~Knopp, ``Viscosity
  quantification using multi-contrast magnetic particle imaging,'' \emph{New
  Journal of Physics}, vol.~20, no.~8, p. 83001, aug 2018.

\bibitem{moddel2020estimating}
M.~Möddel, F.~Griese, T.~Kluth, and T.~Knopp,
  ``\BIBforeignlanguage{en}{Estimating orientation using multi-contrast mpi},''
  \emph{\BIBforeignlanguage{en}{International Journal on Magnetic Particle
  Imaging}}, vol.~6, no.~2, pp. 1--3, 2020.

\bibitem{Viereck2017}
\BIBentryALTinterwordspacing
T.~Viereck, C.~Kuhlmann, S.~Draack, M.~Schilling, and F.~Ludwig,
  ``Dual-frequency magnetic particle imaging of the brownian particle
  contribution,'' \emph{Journal of Magnetism and Magnetic Materials}, vol. 427,
  pp. 156--161, 2017. [Online]. Available:
  \url{http://www.sciencedirect.com/science/article/pii/S0304885316329031}
\BIBentrySTDinterwordspacing

\bibitem{Tay_relaxation_wall_2017}
\BIBentryALTinterwordspacing
Z.~W. Tay, D.~W. Hensley, E.~C. Vreeland, B.~Zheng, and S.~M. Conolly, ``The
  relaxation wall: experimental limits to improving {MPI} spatial resolution by
  increasing nanoparticle core size,'' \emph{Biomedical Physics {\&}
  Engineering Express}, vol.~3, no.~3, p. 035003, may 2017. [Online].
  Available: \url{https://doi.org/10.1088/2057-1976/aa6ab6}
\BIBentrySTDinterwordspacing

\bibitem{tay2019pulsed}
Z.~W. Tay, D.~Hensley, J.~Ma, P.~Chandrasekharan, B.~Zheng, P.~Goodwill, and
  S.~Conolly, ``Pulsed excitation in magnetic particle imaging,'' \emph{{IEEE}
  Transactions on Medical Imaging}, vol.~38, no.~10, pp. 2389--2399, oct 2019.

\bibitem{tay2016}
Z.~W. Tay, P.~W. Goodwill, D.~W. Hensley, L.~A. Taylor, B.~Zheng, and S.~M.
  Conolly, ``A high-throughput, arbitrary-waveform, mpi spectrometer and
  relaxometer for comprehensive magnetic particle optimization and
  characterization,'' \emph{Scientific Reports}, vol.~6, 2016.

\bibitem{Pantke2019MultiFreq}
\BIBentryALTinterwordspacing
D.~Pantke, N.~Holle, A.~Mogarkar, M.~Straub, and V.~Schulz, ``Multifrequency
  magnetic particle imaging enabled by a combined passive and active drive
  field feed-through compensation approach,'' \emph{Medical Physics}, vol.~46,
  no.~9, pp. 4077--4086, Jul. 2019. [Online]. Available:
  \url{https://doi.org/10.1002/mp.13650}
\BIBentrySTDinterwordspacing

\bibitem{Top2020AWMPS}
C.~Top, ``An arbitrary waveform magnetic nanoparticle relaxometer with an
  asymmetrical three-section gradiometric receive coil,'' \emph{Turkish Journal
  of Electrical Engineering and Computer Sciences}, pp. 1344--1354, 05 2020.

\bibitem{vonGladiss2017}
A.~{Von Gladiss}, M.~Graeser, P.~Szwargulski, T.~Knopp, and T.~M. Buzug,
  ``Hybrid system calibration for multidimensional magnetic particle imaging,''
  \emph{Physics in Medicine and Biology}, vol.~62, no.~9, pp. 3392--3406, 2017.

\bibitem{Gladiss2020a}
A.~von Gladiss, M.~Graeser, A.~Behrends, X.~Chen, and T.~M. Buzug, ``Efficient
  hybrid 3d system calibration for magnetic~particle~imaging systems using a
  dedicated device,'' \emph{Scientific Reports}, vol.~10, no.~1, pp. 1--12, oct
  2020.

\bibitem{Knopp2010_modelbased}
T.~Knopp, T.~F. Sattel, S.~Biederer, J.~Rahmer, J.~Weizenecker, B.~Gleich,
  J.~Borgert, and T.~M. Buzug, ``Model-based reconstruction for magnetic
  particle imaging,'' \emph{IEEE Transactions on Medical Imaging}, vol.~29,
  no.~1, pp. 12--18, 2010.

\bibitem{Graeser2017SR}
M.~Graeser, T.~Knopp, P.~Szwargulski, T.~Friedrich, A.~{Von Gladiss}, M.~Kaul,
  K.~M. Krishnan, H.~Ittrich, G.~Adam, and T.~M. Buzug, ``Towards picogram
  detection of superparamagnetic iron-oxide particles using a gradiometric
  receive coil,'' \emph{Scientific Reports}, vol.~7, no.~1, p. 6872, 2017.

\bibitem{knopp2008trajectory}
T.~Knopp, S.~Biederer, T.~Sattel, J.~Weizenecker, B.~Gleich, J.~Borgert, and
  T.~M. Buzug, ``Trajectory analysis for magnetic particle imaging,''
  \emph{Physics in Medicine and Biology}, vol.~54, no.~2, pp. 385--397, dec
  2008.

\bibitem{Knopp2011d}
T.~Knopp, M.~Erbe, T.~F. Sattel, S.~Biederer, and T.~M. Buzug, ``A fourier
  slice theorem for magnetic particle imaging using a field-free line,''
  \emph{Inverse Problems}, vol.~27, no.~9, p. 95004, 2011.

\bibitem{Lu2013}
K.~Lu, P.~W. Goodwill, E.~U. Saritas, B.~Zheng, and S.~M. Conolly, ``Linearity
  and shift invariance for quantitative magnetic particle imaging,'' \emph{IEEE
  Transactions on Medical Imaging}, vol.~32, no.~9, pp. 1565--1575, 2013.

\bibitem{Kaczmarz1937}
S.~Kaczmarz, ``Angen{\"{a}}herte {A}ufl{\"{o}}sung von {S}ystemen linearer
  {G}leichungen,'' \emph{Bulletin of the International Academy Polonica
  Sciences Letters A}, vol.~35, pp. 355--357, 1937.

\bibitem{Boberg_2021}
\BIBentryALTinterwordspacing
M.~Boberg, N.~Gdaniec, P.~Szwargulski, F.~Werner, M.~Möddel, and T.~Knopp,
  ``Simultaneous imaging of widely differing particle concentrations in {MPI}:
  problem statement and algorithmic proposal for improvement,'' \emph{Physics
  in Medicine {\&} Biology}, vol.~66, no.~9, p. 095004, apr 2021. [Online].
  Available: \url{https://doi.org/10.1088/1361-6560/abf202}
\BIBentrySTDinterwordspacing

\bibitem{Knopp2011c}
T.~Knopp, S.~Biederer, T.~F. Sattel, M.~Erbe, and T.~M. Buzug, ``Prediction of
  the spatial resolution of magnetic particle imaging using the modulation
  transfer function of the imaging process,'' \emph{IEEE Transactions on
  Medical Imaging}, vol.~30, no.~6, pp. 1284--1292, 2011.

\bibitem{weber2015enhancement}
A.~Weber, J.~Weizenecker, U.~Heinen, M.~Heidenreich, and T.~M. Buzug,
  ``Reconstruction enhancement by denoising the magnetic particle imaging
  system matrix using frequency domain filter,'' \emph{IEEE Transactions on
  Magnetics}, vol.~51, no.~2, pp. 1--5, 2015.

\bibitem{bezanson2017julia}
J.~Bezanson, A.~Edelman, S.~Karpinski, and V.~B. Shah, ``Julia: A fresh
  approach to numerical computing,'' \emph{SIAM Review}, vol.~59, no.~1, pp.
  65--98, 2017.

\bibitem{Knopp2016}
\BIBentryALTinterwordspacing
T.~Knopp, T.~Viereck, G.~Bringout, M.~Ahlborg, A.~von Gladiss, C.~Kaethner,
  A.~Neumann, P.~Vogel, J.~Rahmer, and M.~M{\"{o}}ddel, ``{MDF: Magnetic
  Particle Imaging Data Format},'' \emph{arXiv preprint arXiv:1602.06072},
  p.~9, 2016. [Online]. Available: \url{http://arxiv.org/abs/1602.06072}
\BIBentrySTDinterwordspacing

\bibitem{knopp2019mpireco}
T.~Knopp, P.~Szwargulski, F.~Griese, M.~Grosser, M.~Boberg, and
  M.~M{\"{o}}ddel, ``{MPIReco.jl}: Julia package for image reconstruction in
  {MPI},'' \emph{International Journal on Magnetic Particle Imaging}, vol.~5,
  no.~1, 2019.

\bibitem{MPIReco}
``{MPIReco.jl: Julia package for image reconstruction in MPI},''
  \url{https://github.com/MagneticParticleImaging/MPIReco.jl.git}, accessed:
  2021-01-10.

\bibitem{Shasha2019NanoparticleCS}
C.~Shasha, E.~Teeman, and K.~M. Krishnan, ``Nanoparticle core size optimization
  for magnetic particle imaging,'' \emph{Biomedical Physics \& Engineering
  Express}, 2019.

\bibitem{Ferguson2009_coresize}
\BIBentryALTinterwordspacing
R.~M. Ferguson, K.~R. Minard, and K.~M. Krishnan, ``Optimization of
  nanoparticle core size for magnetic particle imaging,'' \emph{Journal of
  Magnetism and Magnetic Materials}, vol. 321, no.~10, pp. 1548--1551, May
  2009. [Online]. Available: \url{https://doi.org/10.1016/j.jmmm.2009.02.083}
\BIBentrySTDinterwordspacing

\bibitem{weizenecker_mirco-magnetic_2012}
J.~Weizenecker, B.~Gleich, J.~Rahmer, and J.~Borgert, ``Micro-magnetic
  simulation study on the magnetic particle imaging performance of anisotropic
  mono-domain particles,'' \emph{Physics in medicine and biology}, vol.~57, pp.
  7317--7327, 10 2012.

\bibitem{Goodwill2010}
P.~W. Goodwill and S.~M. Conolly, ``The x-space formulation of the magnetic
  particle imaging process: 1-d signal, resolution, bandwidth, snr, sar, and
  magnetostimulation,'' \emph{IEEE Transactions on Medical Imaging}, vol.~29,
  no.~11, pp. 1851--1859, 2010.

\end{thebibliography}

\end{multicols}
\end{document}